\documentclass[12pt,preprint]{aastex}

\shorttitle{Zeta Tau}
\shortauthors{Schaefer et al.}

\begin{document}


\title{Multi-epoch Near-Infrared Interferometry of the Spatially Resolved Disk Around the Be Star $\zeta$ Tau}


\author{G. H. Schaefer\altaffilmark{1}, D. R. Gies\altaffilmark{2}, J. D. Monnier\altaffilmark{3},  N. D. Richardson\altaffilmark{2}, Y. Touhami\altaffilmark{2}, M. Zhao\altaffilmark{4}, \\  X. Che\altaffilmark{3}, E. Pedretti\altaffilmark{5}, N. Thureau\altaffilmark{5}, 
T. ten Brummelaar\altaffilmark{1}, H. A. McAlister\altaffilmark{2}, S. T. Ridgway\altaffilmark{6}, \\ J. Sturmann\altaffilmark{1}, L. Sturmann\altaffilmark{1}, N. H. Turner\altaffilmark{1}, C. D. Farrington\altaffilmark{1}, and P. J. Goldfinger\altaffilmark{1}}

\altaffiltext{1}{The CHARA Array of Georgia State University, Mount Wilson Observatory, Mount Wilson, CA 91023, U.S.A. (schaefer@chara-array.org.)}
\altaffiltext{2}{Center for High Angular Resolution Astronomy and Department of Physics and Astronomy, Georgia State University, P. O. Box 4106, Atlanta, GA 30302-4106; gies@chara.gsu.edu}
\altaffiltext{3}{University of Michigan, Ann Arbor, MI.}
\altaffiltext{4}{Jet Propulsion Laboratory}
\altaffiltext{5}{University of St. Andrews, Scotland, UK.}
\altaffiltext{6}{National Optical Astronomy Observatory, Tucson, AZ.}





\begin{abstract}
We present interferometric observations of the Be star $\zeta$ Tau obtained using the MIRC beam combiner at the CHARA Array.  We resolved the disk during four epochs in 2007--2009.  We fit the data with a geometric model to characterize the circumstellar disk as a skewed elliptical Gaussian and the central Be star as a uniform disk.  The visibilities reveal a nearly edge-on disk with a FWHM major axis of $\sim$ 1.8 mas in the $H$-band.  The non-zero closure phases indicate an asymmetry within the disk. Interestingly, when combining our results with previously published interferometric observations of $\zeta$ Tau, we find a correlation between the position angle of the disk and the spectroscopic $V/R$ ratio, suggesting that the tilt of the disk is precessing.  This work is part of a multi-year monitoring campaign to investigate the development and outward motion of asymmetric structures in the disks of Be stars.  

\end{abstract}





\setcounter{footnote}{6}

\section{Introduction}

$\zeta$ Tau (HR 1910, HD 37202, HIP 26451) is a bright Be star with an extensive observational history that spans photometric, spectroscopic, polarimetric, and interferometric techniques.  The spectrum of $\zeta$~Tau shows the standard double-peaked H$\alpha$ profile, indicative of a disk in Keplerian rotation \citep[e.g.][]{porter03}.  The relative height of the blue- and red-shifted peaks of the emission lines ($V/R$ ratio) shows cyclic variability measured on the order of $\sim$~1429 days \citep{rivinius06,pollmann08,ruzdjak09,stefl09}.  The spectra of $\zeta$ Tau also occasionally show more complex triple-peaked and ``shell'' profiles.  Current models suggest that the line profile variations in $\zeta$~Tau and other Be stars can be explained by global one-armed oscillation models \citep[e.g.][]{carciofi09}.

$\zeta$ Tau is also a single-lined spectroscopic binary with a period of 133 days \citep[e.g.][and references therein]{harmanec84,ruzdjak09}.  The companion has not yet been detected directly.  By placing limits on the orbital inclination, the mass function indicates that the companion is a low-mass object\citep[$\sim$~1~$M_\odot$;][]{ruzdjak09}, several magnitudes fainter than the primary Be star.  The companion could be a main sequence star, a neutron star, a white dwarf, or an evolved hot subdwarf like that found in $\phi$~Per \citep{gies98}.  However, as \citet{ruzdjak09} discuss, a hot subdwarf may heat the outer facing regions of the disk, producing narrow emission lines like \ion{He}{1} $\lambda 6678$.  The absence of such emission in the spectrum of $\zeta$~Tau suggests that the companion is not a strong flux source. \citet{floquet89} rule out a cool luminous giant based on measurements of the infrared flux.  

Given its brightness ($V_{\rm mean}$=3.0, $H_{\rm mean}$=3.0) and distance \citep[$d$=128 pc;][]{hipparcos}, $\zeta$~Tau is an ideal source for studying the structure and dynamics of its circumstellar disk using optical/infrared (IR) interferometry.  Early spatially resolved measurements revealed the elliptical shape of the $\zeta$~Tau disk \citep{quirrenbach94,quirrenbach97,baldwin02,tycner04}, indicating that the circumstellar material is contained in a flattened disk inclined nearly edge-on to the line of sight.  \citet{gies07} determined the geometry and density structure by fitting an isothermal model of a disk in Keplerian rotation to CHARA Classic interferometric data of the $K$-band emission.  Using the GI2T interferometer, \citet{vakili98} detected an asymmetry in the disk of $\zeta$~Tau based on differential phases measured across the H$\alpha$ line.  These results suggest a shift in the position of a bulge located in the disk, consistent with the prograde motion of a one-armed spiral oscillation in the disk.  Moreover, the nature of the asymmetry and its relation to the temporal $V/R$ profile variations were investigated further by \citet{stefl09} and \citet{carciofi09} using VLTI/AMBER observations.  The visibilities and differential phases measured across the Br$\gamma$ line are consistent with an oscillation pattern created by a one-armed spiral in the disk.

In this paper we present multi-epoch interferometric observations of $\zeta$~Tau obtained using the CHARA Array in the $H$-band.  The visibilities provide information about the size and orientation of the disk while the closure phases indicate the presence of an asymmetry within the light distribution.  We fit the data from each epoch using a model consisting of the central star and a skewed elliptical Gaussian disk.  We present the results in \S~\ref{sect.model}.  In \S\ref{sect.disc}, we describe the physical characteristics of the model and discuss the changes we observe in the orientation of the disk on the sky, the asymmetric light distribution, and their relation to the cyclic variations measured in the emission line profiles.  In \S\ref{sect.precess}, we outline the details of a precession model that can explain the changes we observe in the disk of $\zeta$~Tau.  We summarize the results of the study in \S\ref{sect.summary}.

\section{CHARA Array Observations}
\label{sect.obs}

The CHARA Array is an optical/IR interferometer located on Mount Wilson \citep{tenbrummelaar05}.   The array has six 1-meter telescopes arranged in a Y-configuration with baselines ranging from 34--331 meters.  We used the Michigan Infrared Combiner \citep[MIRC;][]{monnier04,monnier06a} at the CHARA Array to observe the disk of $\zeta$ Tau in the $H$-band.  MIRC combines the light from four telescopes simultaneously, providing visibility amplitudes on six baselines and closure phases on four triangles.  It uses single-mode optical fibers to spatially filter the light.  The fibers are brought together by a V-groove array in a nonredundant pattern which encodes the overlapping fringes formed from the outgoing light with distinct, spatial interference frequencies.  We used the low spectral resolution prism (R$\sim$50) to disperse the fringes across eight spectral channels in the $H$-band ($\lambda$~=~1.5-1.8~$\mu$m).

Table~\ref{tab.log} provides a log of the observations that lists the UT date of the observation, the configuration of telescopes used, and the observed calibrator stars.  On 2007 Nov 11--14 we used the inner array (S2-E2-W1-W2) of shorter baselines ranging from 108--248 m.  
On the remaining dates we used the outer array (S1-E1-W1-W2) of longer baselines ranging from 108--331 m.  To calibrate the interferometric observations, we also observed single stars with angular diameters smaller than 0.9 mas 
\citep[selected from the catalogs of][]{pasinetti01,merand05,richichi05,vanbelle08}.  
We derived angular diameters of the calibrators by fitting all the available flux data 
with reddened models of the spectral energy distribution. 
In the best cases, we included UV-fluxes from the {\it International
Ultraviolet Explorer} satellite, optical spectrophotometry,
and near-IR fluxes based upon 2MASS and other sources. 
The more poorly studied stars have flux estimates at least 
for the Johnson $UBV$ bands and for the $JHK_s$ 2MASS bands. 
The final ingredient in the fit is an estimate of the stellar
effective temperature $T_{\rm eff}$ that usually comes from detailed 
spectroscopic investigations.  The observed spectral 
energy distributions are then fit with low resolution 
spectra from the models of 
R.\ Kurucz\footnote{http://kurucz.harvard.edu/grids.html}
(for $T_{\rm eff} > 5000$ K)
or Gustafsson et al.\ (2008; for $T_{\rm eff} < 5000$ K). 
Table~\ref{tab.cal} lists the names and HD numbers of the calibrators, spectral classification, $V$ and $H$ magnitudes, the adopted $T_{\rm eff}$ and reference source 
(also for the adopted value of gravity $\log g$), 
as well as the derived interstellar reddening \citep[usually for a ratio 
of total to selective extinction, $R=3.1$;][]{fitzpatrick99}, 
and the limb darkened angular diameter.  In what follows we assume for simplicity 
that the limb darkened angular diameter equals the uniform 
disk angular diameter (since the differences are very small 
in the near-IR where limb darkening is minimal).  

The data were reduced using the standard MIRC reduction pipeline \citep[e.g.][]{monnier07b}.  The reduction process involves Fourier transforming the raw background-subtracted data to obtain fringe amplitudes and phases.  Corrections are then applied to the visibility amplitudes to account for the fiber coupling efficiencies to correct for the different flux levels in each of the four beams.  The data from 2009 Nov 10 were acquired using the photometric channels recently installed in MIRC that divert 20\% of the light going through each fiber to measure directly the contribution of light from each telescope \citep{monnier08,che10}.  Drifts in the overall system response are calibrated by using single stars of known sizes observed before and after the target observations.  The reduction pipeline outputs calibrated squared visibilities and closure phases in the OIFITS format\footnote{Calibrated OIFITS data are available upon request.} \citep{pauls05}.  The errors in these values are derived by combining the scatter in the measured data with the errors propagated through the calibration process.

\begin{figure}
   \scalebox{0.9}{\includegraphics{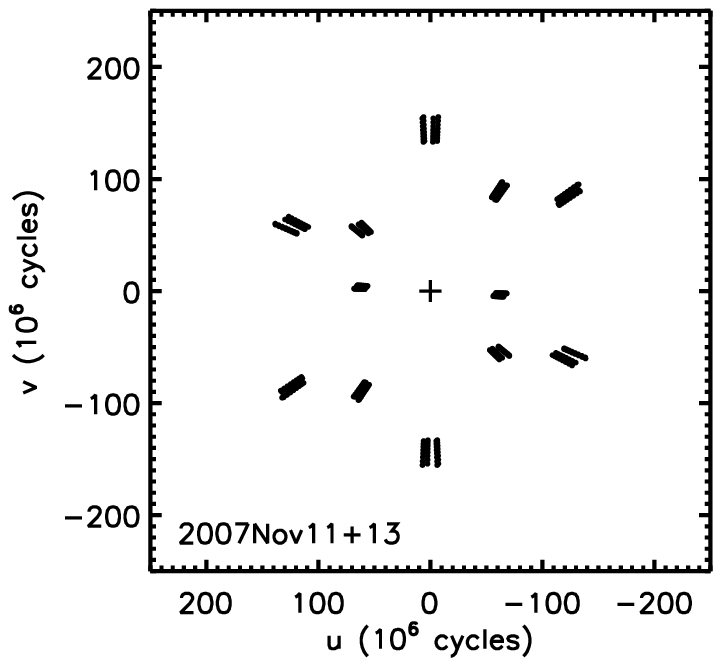}} 
   \scalebox{0.9}{\includegraphics{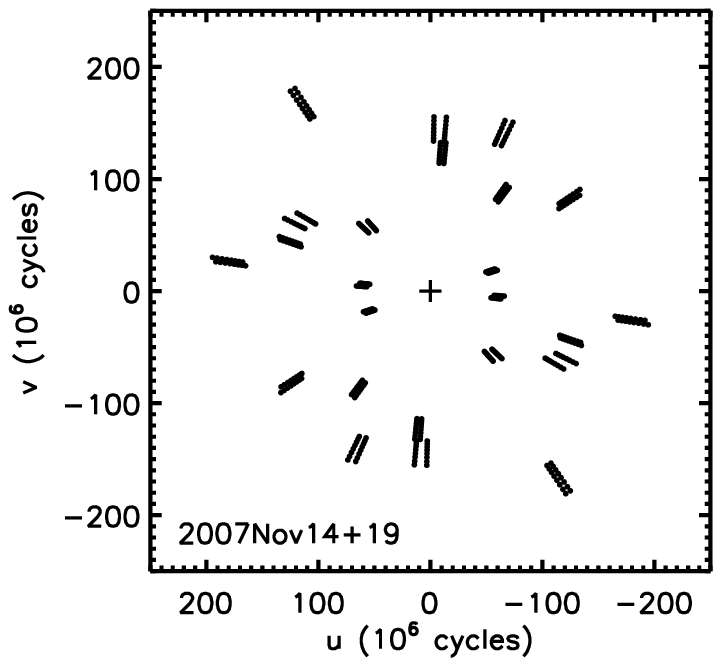}}  \\
   \scalebox{0.9}{\includegraphics{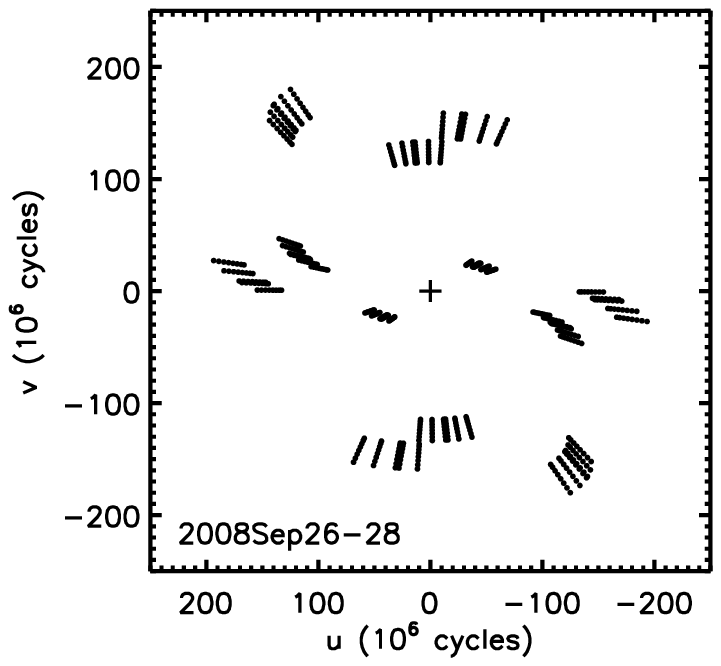}}
   \scalebox{0.9}{\includegraphics{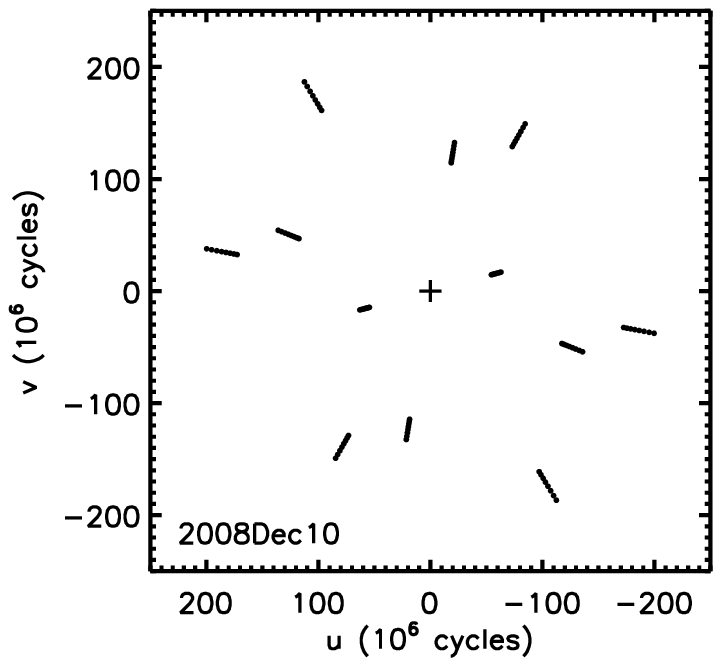}} \\
   \scalebox{0.9}{\includegraphics{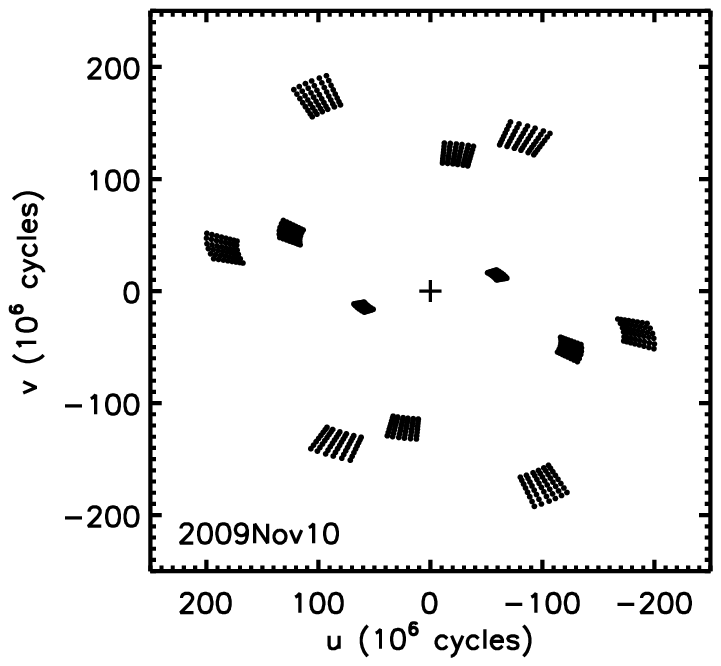}}
   \caption{$u$--$v$ coverage on the sky during the MIRC observations of $\zeta$ Tau in 2007--2009.  }
\label{fig.uv}
\end{figure}

\begin{figure}
   \scalebox{0.40}{\includegraphics{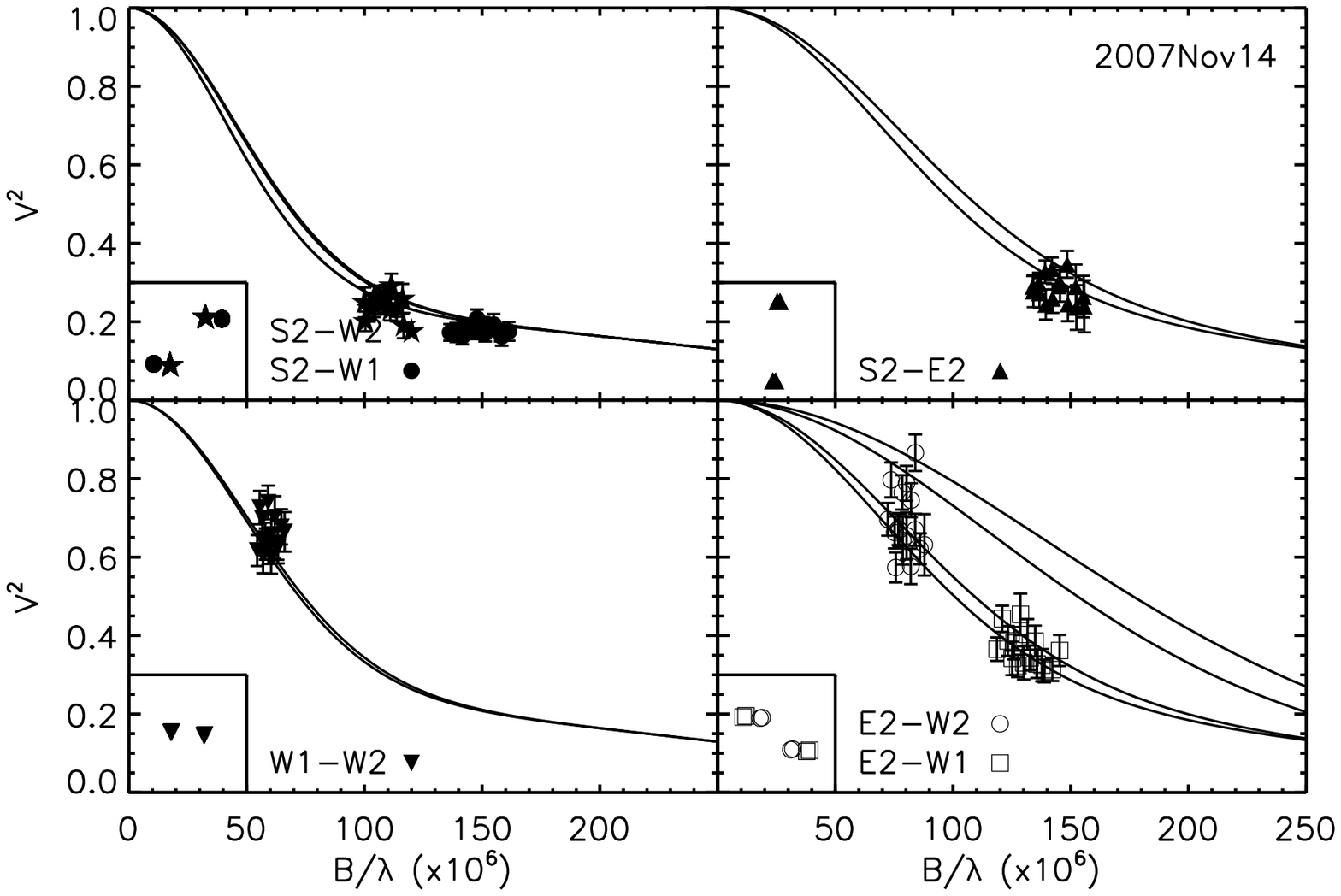}}
   \scalebox{0.40}{\includegraphics{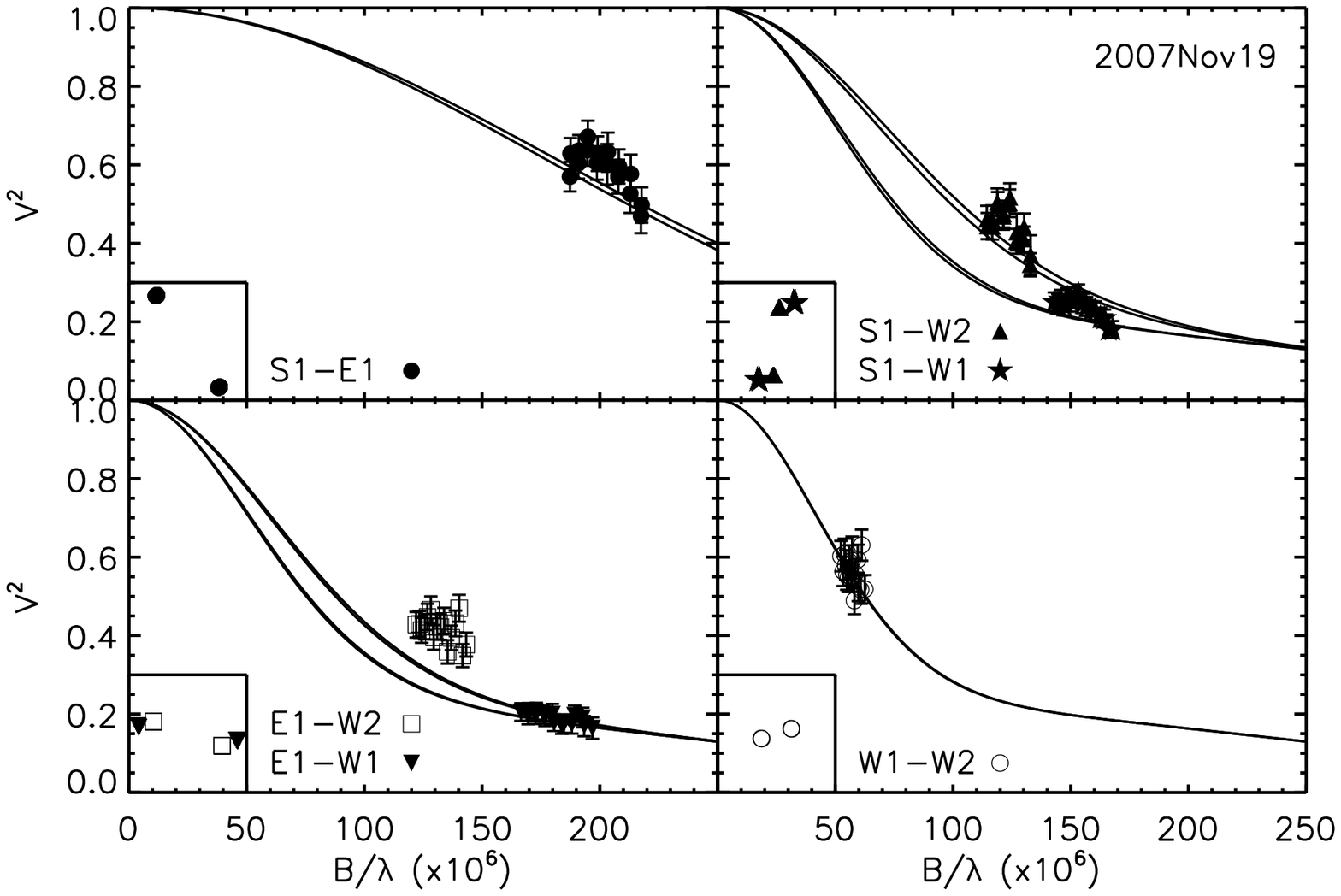}} \\
   \scalebox{0.40}{\includegraphics{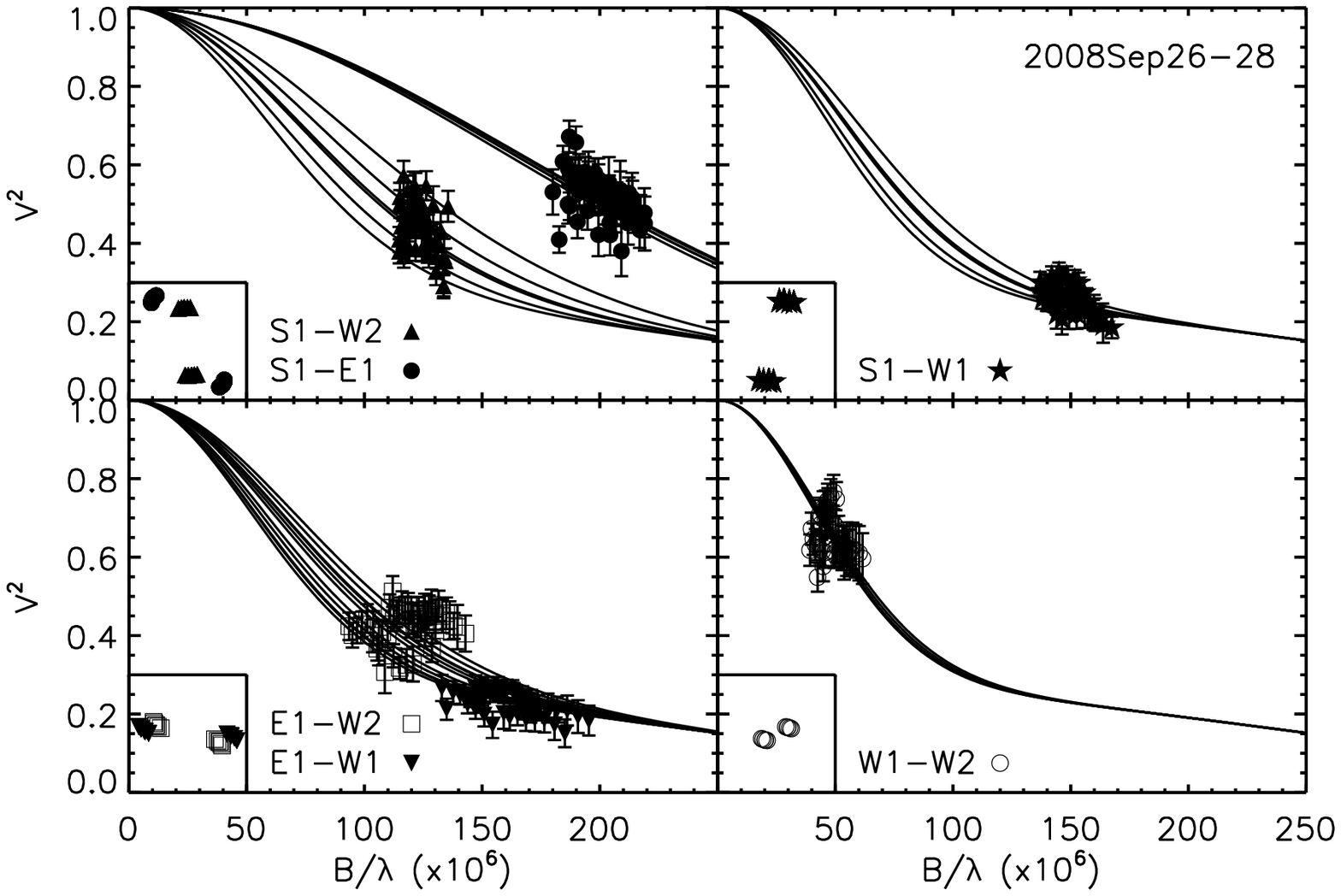}}
   \scalebox{0.40}{\includegraphics{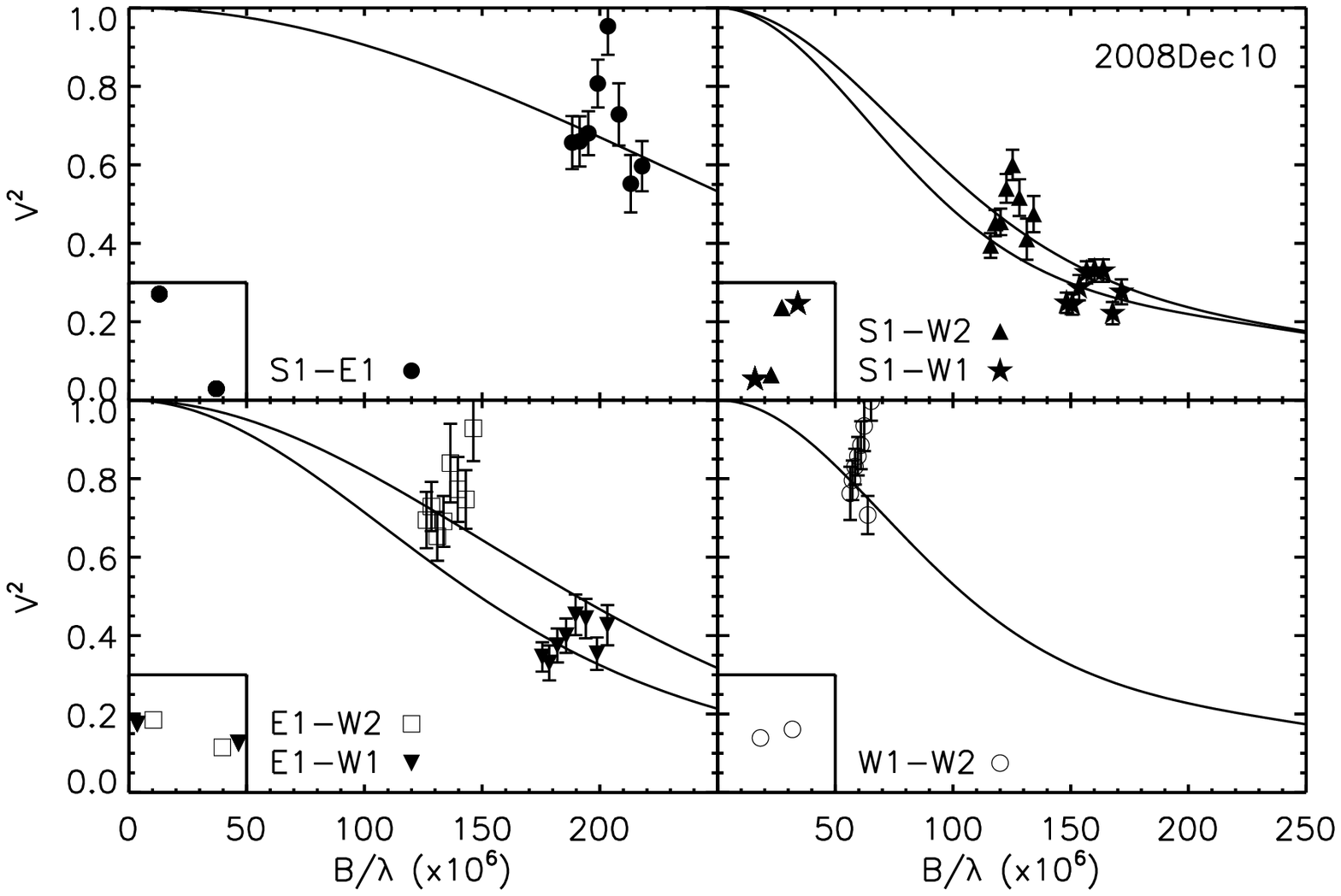}} \\
   \scalebox{0.40}{\includegraphics{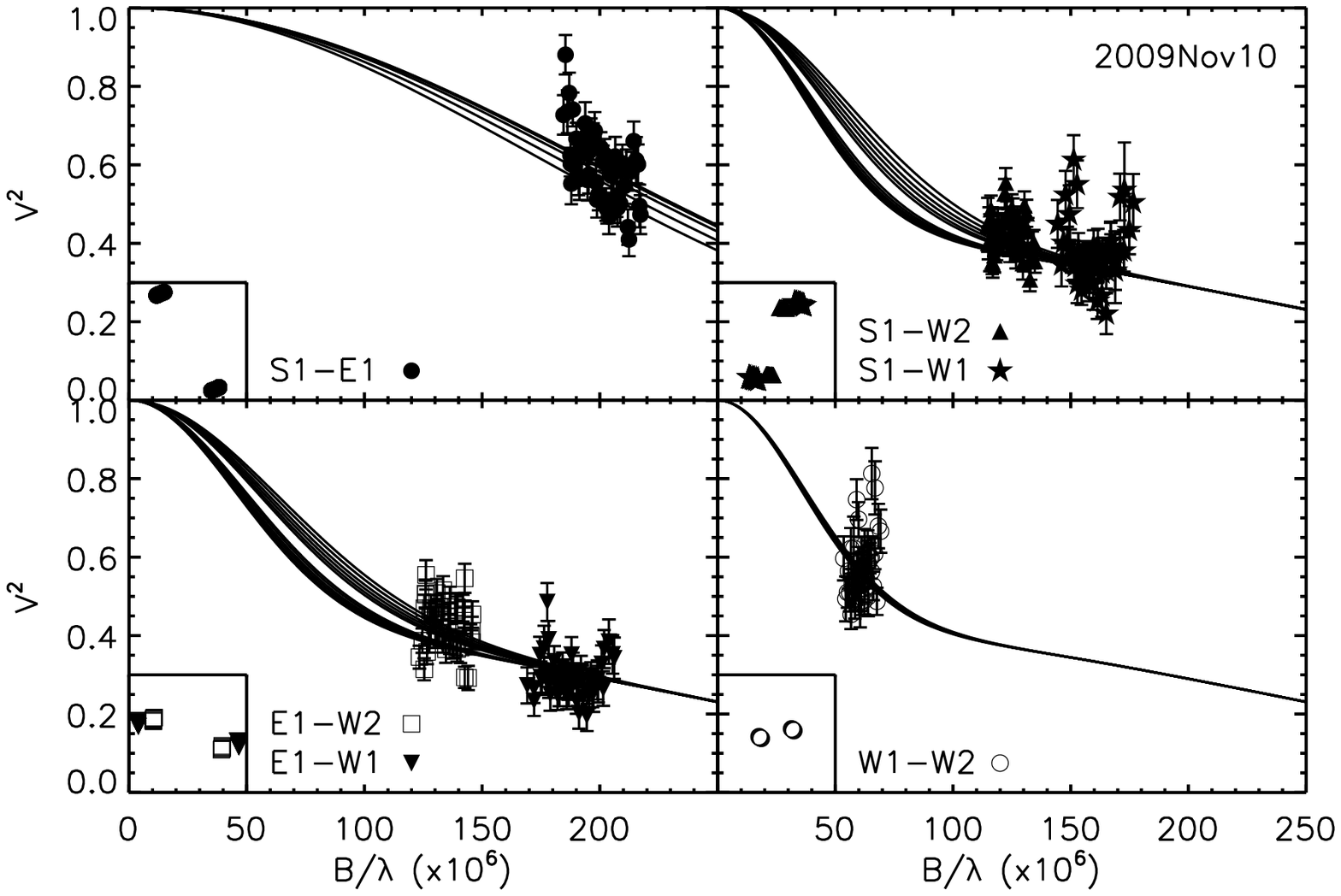}}
   \caption{Squared visibilities measured for $\zeta$~Tau using MIRC in 2007--2009.  For clarity, the measurements during each epoch are grouped by proximity in position angle of the baseline.  The small inset panels show the projection of observed $u$--$v$ points on the plane of the sky.  The solid lines show the best global-epoch fits given in Table~\ref{tab.diskfit}.  There are a number of lines in each plot to show the model at each of the observed $u$--$v$ projections.}
\label{fig.vis2}
\end{figure}

\begin{figure}
   \scalebox{0.65}{\includegraphics{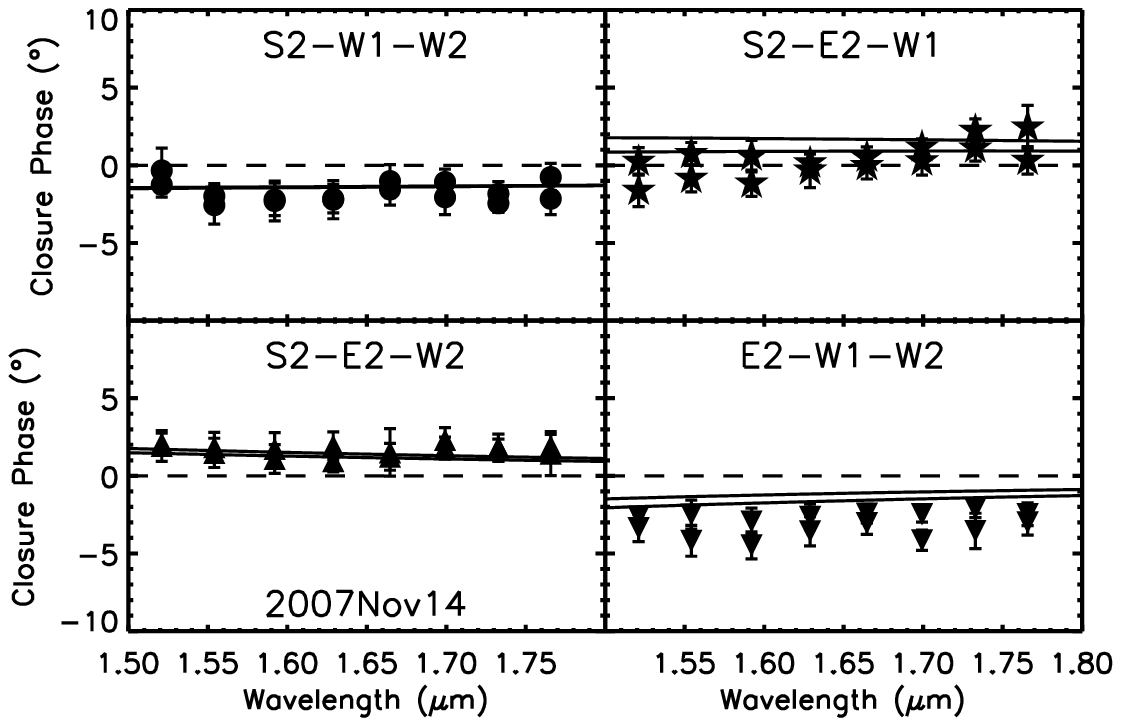}} 
   \scalebox{0.65}{\includegraphics{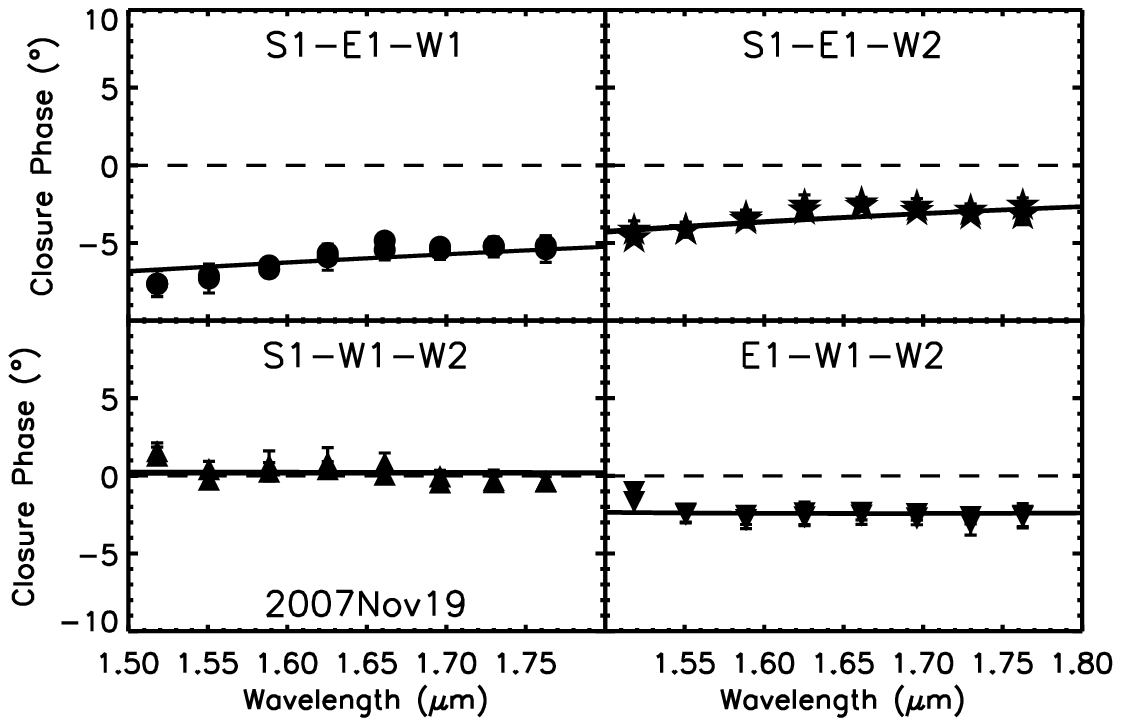}} \\
   \scalebox{0.65}{\includegraphics{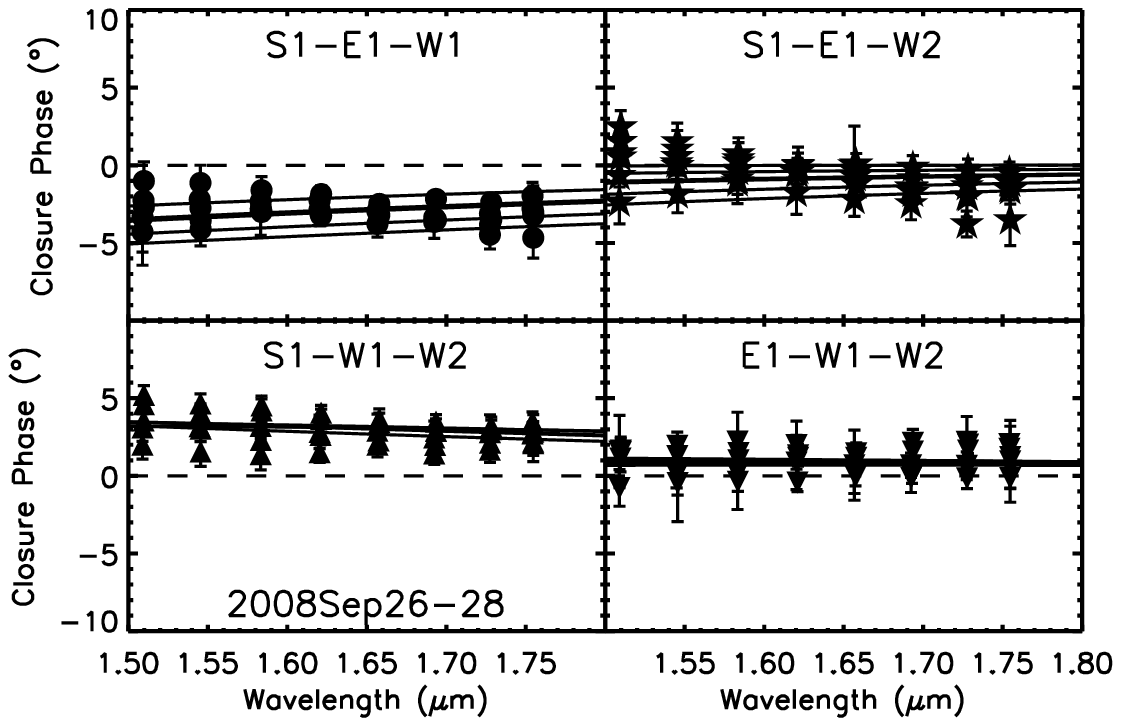}} 
   \scalebox{0.65}{\includegraphics{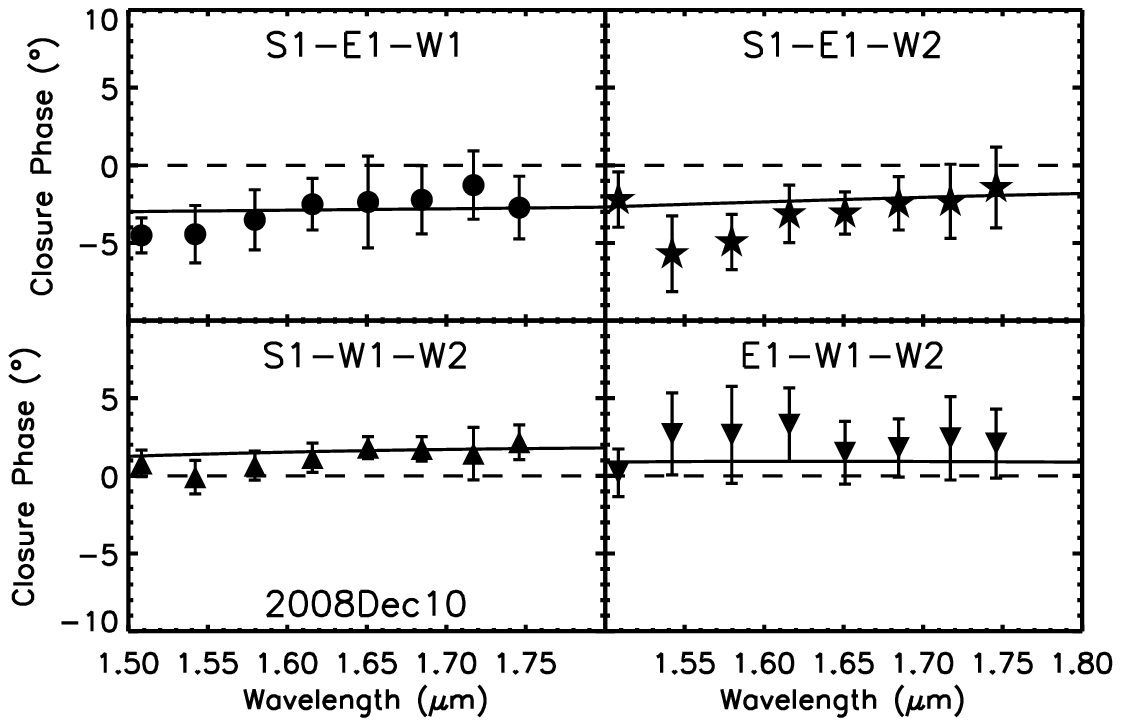}} \\
   \scalebox{0.65}{\includegraphics{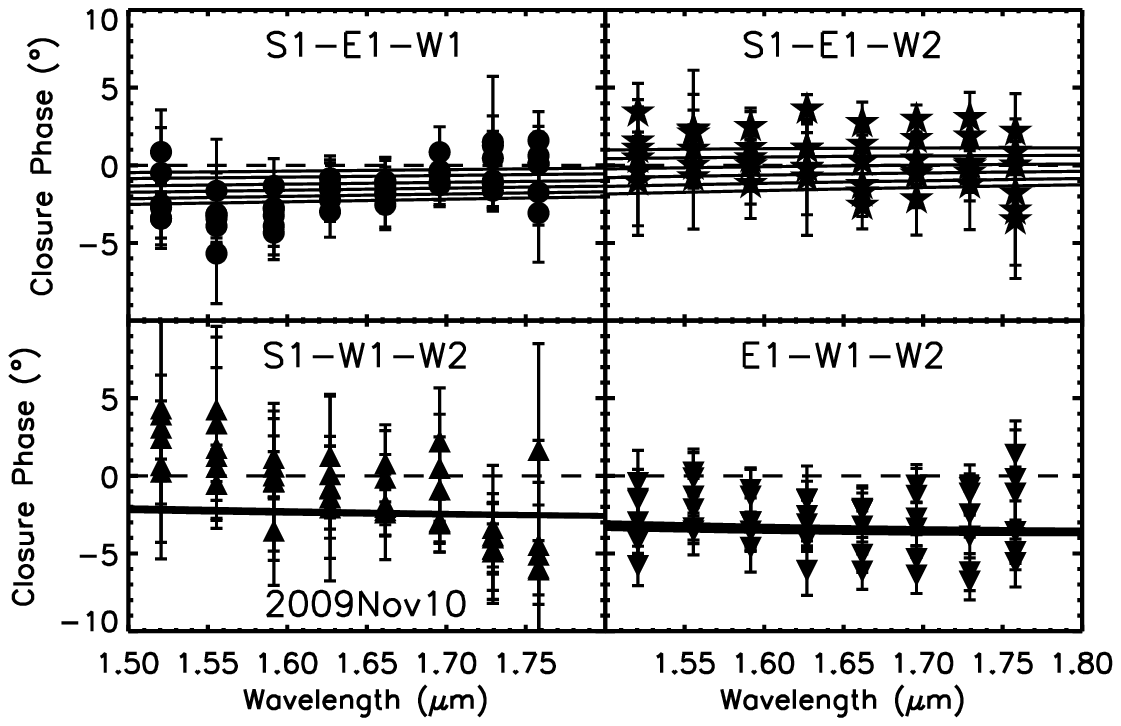}} 
   \caption{Closure phases measured on the four closed triangles during our MIRC observations in 2007--2009.  The solid lines show the best global-epoch fits given in Table~\ref{tab.diskfit}.  The multiple lines show how the model changes across the observed $u$--$v$ projections.}
\label{fig.t3}
\end{figure}

Figure~\ref{fig.uv} shows the $u$--$v$ coverage on the sky sampled by the CHARA Array during the epochs of the $\zeta$~Tau observations obtained with MIRC.  The squared visibilities measured on the six baselines are shown in Figure~\ref{fig.vis2}.  The range of angular sizes indicated by the visibilites along different projection angles reflects the elliptical shape of the $\zeta$ Tau disk on the plane of the sky.  The closure phases on the four closed triangles are shown in Figure~\ref{fig.t3}.  The non-zero closure phases indicate the presence of an asymmetry in the disk.




\section{Modeling the Disk of $\zeta$ Tau}
\label{sect.model}

We fit a two-component geometric model to the MIRC data obtained for $\zeta$~Tau.  The model is composed of a uniform disk with an angular diameter of 0.40~mas \citep[$R$=5.5 $R_{\odot}$;][]{gies07} to fit the central star and an elliptical, Gaussian surface brightness distribution to model the circumstellar disk.  To account for the asymmetry we detect in the closure phases, we modulated the elliptical Gaussian disk by a sinusoid as a function of projected azimuth or position angle \citep[e.g.][]{monnier06b,thureau09}.  This creates a ``skewed'' disk model where the sinusoid causes the brightness distribution to peak on one side of the disk and places a depression in the brightness on the other side.  
The intensity distribution on the sky of the asymmetric, elliptical, Gaussian disk is given by the following functional form,
\begin{equation}
y' = x\sin{\phi_{\rm maj}} + y\cos{\phi_{\rm maj}}
\end{equation}
\begin{equation}
x' = x\cos{\phi_{\rm maj}} - y\sin{\phi_{\rm maj}} 
\end{equation}
\begin{equation}
I_{\rm disk} = I_0 ~[1 + {A_{\rm skew}} * \cos^p{(\phi_{\rm skew} - \phi)}]\exp \left\{ -4 \ln{2} \left[ \left(\frac{x'}{\theta_{\rm min}}\right)^2 + \left(\frac{y'}{\theta_{\rm maj}}\right)^2 \right] \right\}
\end{equation}
where $\theta_{\rm maj}$ and $\theta_{\rm min}$ are the full-width at half maximum of the major and minor axes of the Gaussian disk, $\phi_{\rm maj}$ is the position angle of the major axis measured east of north, $A_{\rm skew}$ is the amplitude of the sinusoidal modulation (0$-$1), $\phi_{\rm skew}$ is the position angle of the skew maximum intensity measured east of north, and $I_0$ is the normalized central brightness of the disk.  In this parametrization, $+$$x$ runs in the direction of positive RA (east), $+$$y$ in the direction of positive declination (north), and $\phi = \arctan(x/y)$.  The shape of the asymmetry can be flattened out to be more boxy in appearance by raising the sinusoid by the ``skew power'' $p$ ($p < 1.0$).  To avoid taking the root of a negative number, we take the absolute value of $\cos{(\phi_{\rm skew} - \phi)}$, raise it to the power $p$, and then multiply that value by the original sign of $\cos{(\phi_{\rm skew} - \phi)}$.  Finally, we scale the star and disk contributions by the fraction of their $H$-band fluxes ($f_{\rm star}$ and $f_{\rm disk}$).

To fit the parameters of the $\zeta$ Tau model we performed a Levenberg-Marquardt least-squares minimization using the IDL mpfit\footnote{http://cow.physics.wisc.edu/\raisebox{0.1em}{\tiny$\sim$\,}craigm/idl/idl.html} rountine developed C. B. Markwardt.  For each iteration, we computed an image of the geometric model and computed the visibilities and closure phases at the same $u$-$v$ coordinates as the data and compared these with the observed values.   Table~\ref{tab.diskfit} lists the model parameters derived for each epoch.  We present the total reduced $\chi^2_\nu$(all) for a solution and also break it down into separate contributions from the visibilities $\chi^2_\nu(V^2)$ and closure phases $\chi^2_\nu$(T3).  The formal 1\,$\sigma$ uncertainties listed in Table~\ref{tab.diskfit} were computed from the square root of the diagonal elements in the covariance matrix and scaled by $\chi^2_\nu$.  To test the reliability of the derived parameters, we fit the data from each night separately in addition to doing a global fit to data observed on multiple nights at the same epoch.  For the 2007 Nov data, we decided to group only the 2007 Nov 14 (inner array) and 2007 Nov 19 (outer array) data in the global-epoch fit.  We did this to ensure that the number of data points going into the global fit from the inner and outer arrays were roughly equal (to provide equal weights for each configuration).  Additionally, the data from 2007 Nov 14 yielded the best individual fit for the inner array; the data quality from 2007 Nov 11 and 13 was not as good because of poorer seeing conditions.  In Figures~\ref{fig.vis2} and \ref{fig.t3}, we overplot the model in each panel to compare with the measured visibilities and closure phases. Images of these models derived for 2007 Nov 14+19, 2008 Sep 26--28, 2008 Dec 10, and 2009 Nov 10 are shown in Figure~\ref{fig.model}.  
We note that the $\zeta$~Tau data on 2008 Dec 10 were taken during poor seeing conditions that improved by the time we observed the second calibrator of the night.  The changing seeing conditions will result in larger systematic calibration errors.  We noted a trend of increasing system visibility on all baselines during that night, so we suspect that the calibration
will affect the overall scaling of the disk size more than it affects
the orientation or axis ratio of the disk.  Additionally, the 2008 Dec 10 data only had one sampling in $u$-$v$ space, where the other epochs and nights had several observations to improve confidence in the results.

The largest source of systematic error lies in the calibration of the MIRC visibilities.  The main contributions come from the photometric corrections for the amount of light in each fiber and errors in the angular diameters we assumed for the calibrator stars.  We estimate that these effects can result in a $\sim$10\% systematic uncertainty in the squared visibilities.  To investigate how this affects the model parameters, we generated 100 data sets for each epoch where the visibilities on each baseline were altered by a fraction of $\pm$10\%; the values of the offsets were drawn randomly from a Gaussian distribution.  Because the visibilities are correlated across the 8 spectral channels, we varied all of the visibilities on a given baseline by the same fraction.   Using these modified data sets, we derived a new set of best fit parameters for the disk model.  We then estimated the systematic effects on a given parameter by taking the standard deviation of the parameter distribution generated from the modified data sets.  Table~\ref{tab.syserr} shows the size of these effects on the fraction of flux in the disk, the major axis, minor axis, and position angle of the disk.  These parameters are the most sensitive to the visibility calibration.  The size of the systematic effects are typically 2--5 times the formal internal errors determined from the covariance matrix (Table~\ref{tab.diskfit}).  The systematic errors tend to have a greater impact on the data from the inner array than the outer array.  This is because a 10\% change in the visibility calibration will have a larger effect on shorter baselines where the source is less resolved and the visibilities are closer to 1.0.

\section{Discussion}
\label{sect.disc}

\subsection{Size and brightness of the disk in the $H$-band}

Across the epochs with reliable visibility calibration (excluding 2008 Dec 10), the FWHM of the major axis of the disk ranges between 1.6-2.1 mas (0.20-0.27 AU).  The range of variation does not significantly exceed the systematic uncertainties, so we cannot conclusively state whether we measure a true change in the outer radius of the disk.  The size in the $H$-band is similar to the FWHM of the major axis of 1.8 mas measured in the $K'$-band by \citet{gies07} and is smaller than the 3.1-4.5 mas FWHM measured in H$\alpha$ by \citet{quirrenbach97} and \citet{tycner04}.

It is interesting to note that the inner array data in 2007 Nov 11-14 tend to favor a puffier disk, where the ratio between the size of the minor axis relative to the major axis is larger than the ratio derived from the outer array data on 2007 Nov 19.  If this effect is real, it might suggest a polar wind component along the direction of the minor axis \citep[e.g.][]{kanaan08,kervella09} that is less resolved with the shorter baselines of the inner array and more resolved with the longer baselines of the outer array.  We plan to follow-up on this in future observations; the newly installed MIRC photometric channels will also improve the visibility calibration and help to discern whether the effect is real.

In general, a degeneracy exists between the size of a circumstellar disk and the flux ratio of the star-to-disk contributions.  This degeneracy breaks down if we successfully resolve the disk and can see the visibility curve flatten out at the longest baselines, indicating that the disk is fully resolved and hence, the visibility amplitude is dominated by the contribution of the underlying unresolved star.  As seen in Figure~\ref{fig.vis2}, the visibility amplitude begins to flatten out on long baselines that sample projections along the major axis of the $\zeta$ Tau disk, indicating that we are able to remove much of this degeneracy.  From our MIRC observations, we find that the star contributes on average about 0.55 $\pm$ 0.08 of the light in the $H$-band.  In comparison, the star contributes 0.41 of the flux in the $K'$-band \citep{gies07}.  
\citet{touhami10} recently measured the near-IR excess flux in $\zeta$ Tau and other Be stars by assuming that the  stellar flux dominates in the visible part of the spectrum.    The neglect of any disk flux in the visible will result in an overestimate of the stellar flux in the near-IR.  Therefore, their upper limits on the ratio of stellar to total flux of 0.76 in $H$ and 0.62 in $K$ are consistent with the interferometric results.

\subsection{Position Angle Variations and Disk Precession}
\label{sect.precess}

The position angle of the major axis of the $\zeta$ Tau disk changes across the different epochs of MIRC observations.  For comparison, we list the model parameters of prior results from the literature in Table~\ref{tab.compare}; there is some scatter among these position angle measurements also.  We include here a measurement derived from CHARA Classic $K^\prime$-band observations by Y.\ Touhami et al.\ (in preparation) that were made contemporaneously with the MIRC observations in 2008.  These position angle variations probably reflect changes in the disk spatial flux distribution in the sky that are related to disk gas asymmetries.

The disks of Be stars may develop a global, one-armed spiral ($m=1$ mode) instability \citep{kato83,okazaki91,okazaki97,papaloizou92,papaloizou06,ogilvie08,oktariani09}.  The oscillation mode appears as a spiral density wave that precesses prograde with the disk rotation and completes a $360^\circ$ advance over a timescale of a few years.  The progressive change in gas density with disk azimuth is probably the explanation for cyclic changes in the intensities of the violet $V$ and red $R$ peaks of the H$\alpha$ emission profiles.  Recent theoretical work by \citet{ogilvie08} and \citet{oktariani09} suggests that there is significant vertical motion associated with the spiral arm that might assume the form of a tilted disk that precesses with the one-armed oscillation mode.  \citet{hummel98} and \citet{hirata07} both argued that precession of the disk tilt is the cause of long-term emission profile variations in some Be stars.

The cyclic variations in the $V/R$ ratio of the H$\alpha$ emission peak strengths is well documented in the case of $\zeta$ Tau \citep{rivinius06,pollmann08,ruzdjak09,stefl09}.  The evidence relating the $V/R$ variations to the oscillation model is especially striking for the last three cycles when the H$\alpha$ emission peaks varied with a cycle time of 1429 days and with the $V/R$ maximum occurring at a reference epoch of JD 2,450,414 \citep{stefl09,carciofi09}.  The H$\alpha$ variations follow a cycle where the violet side of the emission line peaks at oscillation phase $\tau = 0$ ($V>R$), descends to $V=R$ at $\tau = 0.25$ when a central, ``shell'', absorption component appears, then the red peak reaches a maximum at $\tau = 0.5$ ($V<R$), and ascends back through $V=R$ at $\tau = 0.75$ where a third central emission peak feature appears. \citet{carciofi09} argue that these spectral variations are caused by the changing orientation of a spiral density enhancement around the disk azimuth.  They also present evidence for an asymmetry in the disk based on the astrometric shift in the photocenter of the blue and red-shifted parts of the Br$\gamma$ line measured with VLTI/Amber.

We collected more recent H$\alpha$ spectra to confirm that the $V/R$ variation continued through the time span of the MIRC observations.  The spectra were downloaded from the BeSS database maintained at the GEPI laboratory of the Observatoire de Paris-Meudon\footnote{http://basebe.obspm.fr/basebe/} and they were augmented by two spectra from the University of Toledo Ritter Observatory (courtesy of E.\ Hesselbach and K.\ Bjorkman) and one from the Kitt Peak National Observatory Coude Feed telescope (courtesy of E.\ Grundstrom and V.\ McSwain).  All these spectra have a resolving power of approximately $R=10,000$ or better.  They were transformed to continuum flux normalized versions on a standard heliocentric wavelength grid.  The integrated H$\alpha$ emission equivalent width (measured without correction for any photospheric component) was relatively constant over this time with a mean value of $W_\lambda = -15.2$~\AA , similar to the mean over 1992 -- 2008 of $-15.5$ \AA ~\citep{stefl09}.  We measured a simple $V/R$ estimate as the ratio of the maximum flux excess above the continuum on the blue side compared to that on the red side of the H$\alpha$ emission profile, and the results are shown in Figure~\ref{fig.vr} (together with the times of the MIRC observations).  The $V/R$ variation did indeed continue through this period, but the maximum occurred near JD 2,454,505 $\pm 30$ or approximately 196 d earlier than predicted according to the ephemeris from \citet{carciofi09}.  This trend is consistent with the slow decrease in cycle time noted by \citet{stefl09}.  For the purpose of comparing the disk position angles from the MIRC results with the $V/R$ phase $\tau$, we set $\tau$ using the observed recent time of maximum but with the 1429 d cycle time unchanged.  Note that there is evidence of a shorter, $\sim 70$ d variation in the measurements prior to the $V/R$ maximum, first noted in earlier observations by \citet{pollmann08}, that we will discuss below.  A third, central peak sometimes appears in the H$\alpha$ profile near phase $\tau \sim 0.75$ that might confuse the $V/R$ estimate.  The $V/R$ measurements presented in Figure~5 span the range from $\tau=0.91$ to 0.48, and from a visual inspection of the H$\alpha$ profiles, we only see evidence of a weak, central peak in six spectra obtained between HJD~2,454,433 and 2,454,457 ($\tau = 0.95 - 0.97$).  The stronger $V$ and $R$ peaks are well-separated from the low intensity, third peak in all six of these observations, so we doubt that the third peak has any significant influence on this particular set of $V/R$ measurements.

In Figure~\ref{fig.padisk}, we plot the position angle of the disk measured with MIRC and other published interfometric observations as a function of $V/R$ phase $\tau$.   We list the phase computed for each epoch in the bottom row of Tables~\ref{tab.diskfit} and \ref{tab.compare}.  Note that we do not include in Figure~\ref{fig.padisk} the estimate for 1992 from the work of \citet{quirrenbach97} since the $V/R$ variations were not as well documented then and the cycle time was probably significantly longer \citep{rivinius06,ruzdjak09}.  It appears that the position angle did vary with $\tau$ over the last cycle.  A sinusoidal fit to the variation yields a mean position angle of $<\phi_{\rm maj}> = -58\fdg0 \pm 1\fdg4$, a semiamplitude of $8\fdg1 \pm 1\fdg7$, and an epoch of maximum position angle at $\tau = 0.23 \pm 0.03$.  The residuals of the fit are reduced from a rms = $4\fdg6$ for a simple mean to a rms = $2\fdg2$ for the sinusoidal fit, and an $F$-test indicates that such a reduction would only occur $3\%$ of the time for random errors.  Thus, we suggest that the position angle variations are significant and are probably related to the $V/R$ variation.  We note that the fitted mean position angle from interferometry agrees well with the  mean from linear polarization measurements, $-58\fdg1 \pm 1\fdg2$ \citep{mcdavid99,stefl09}.   

We suspect that the position angle variations result from a tilt of the disk that is associated with the one-armed spiral density enhancement.  A cartoon model for the variation is presented in Figure~\ref{fig.precess} that shows the orientation of the circumstellar disk and the density enhancement over the $V/R$ cycle.  For the sake of simplicity, the tilt is shown as constant at all disk radii (for a planar geometry), although we suspect that in reality the disk tilt is largely confined to radii where the density enhancement is largest.  The geometry of the projected disk in the sky is described by four parameters: the mean position angle of the disk normal $\alpha_0$ measured east from north, the average disk inclination $<i>$ (equal to the inclination of the spin axis of the Be star), the precession cone semiangle $\theta$, and the time variable azimuth of the precession axis $\psi (\tau)$ (measured relative to the line of sight).  In the thin disk approximation, the projected position angle of the disk major axis will vary as $$\phi_{\rm maj} (\tau) = \alpha_0 + 90^\circ -\arctan (\sin\psi(\tau) \tan\theta / \sin <i>)$$ and the projected ratio of the minor to major axes will be related to the time variable inclination by $$\cos i(\tau) = \cos\psi(\tau) \sin\theta \sin <i> + \cos\theta \cos <i>.$$  We assume that the disk tilt follows the prograde precession of the one-armed spiral mode and that the disk approaches us in the north-west sector \citep[based upon the interferometric results;][]{vakili98,stefl09}.   Then, we derive from the sinusoidal fit of the position angle variation, $\alpha_0 = -148\fdg0 \pm 1\fdg4$, $\theta = 8\fdg1 \pm 1\fdg7$, and a temporal relation for the azimuthal precession angle $\psi(\tau) = 360^\circ (\tau + (0.52 \pm 0.03))$. Finally we set $<i>=92\fdg8$ based upon the apparent inclination of $i(\tau =0.57) = 85^\circ$ (angle between the disk normal directed south-west and the line of sight) derived by \citet{carciofi09} from their disk model of the VLTI/Amber observations.  

We next consider the relationship between the disk tilt geometry
and the location of the density enhancement.  We place
the enhancement maximum in the tilted disk plane at an arbitrary distance
from the Be star of $2.5 R_\star$ (Carciofi et al.\ 2009), and it is
marked as
a gray circle in each panel of Figure~7.  Recall that $\tau=0$ is defined
by the time of $V/R$ maximum when the density enhancement
is located in the approaching part of the disk in the plane of the sky.
This relation defines the azimuthal placement of the enhancement.
In the cartoon model, the precession motion brings this enhancement
almost directly in front of the Be star at $\tau = 0.25$, which is
presumably the reason for the strong, shell absorption feature
that appears in the H$\alpha$ profile then.  Half a cycle later
at $\tau = 0.75$ the density enhancement is occulted by the Be star,
but we speculate that the spiral arm extension is large enough that
some high density regions are still visible and contribute to
the appearance of a third emission peak near the center of
the H$\alpha$ profile.  The azimuthal relation we find for
$\psi(\tau)$ places the density enhancement maximum near
the line of the nodes between the tilted disk plane
and equatorial plane of the Be star, and it will be interesting
to see if such a placement will be found in future
three dimensional models for the one-armed oscillation.
Note that for this location and disk inclination angle,
we expect that the photocenter of the enhancement will always be
found along the line of the projected stellar equator (dotted line
in Fig.~7).  Thus, the observed position angle variations result
not from the shifting position of the enhancement but from the
precession of the extended disk whose tilt is presumably
generated by vertical motions associated with the enhancement.
Our simple model also predicts that the
observed ratio of disk minor to major axes will be smallest
around $\tau = 0.28$ and 0.68, which is consistent with our
finding of the smallest ratio in the 2008 Dec 10 data at
$V/R$ phase $\tau = 0.21$.

The fact that $\zeta$ Tau has a binary companion in a 133~d orbit \citep{ruzdjak09} probably means that any tilt of the Be star's disk will be modulated by the tidal force of the binary companion.  Approximately twice each orbit, a tilted disk will experience a tidal torque in the direction of coalignment with the orbital plane, and this results in a nodding motion that is seen, for example, in the precessing disk and jets of massive X-ray binary SS~433 \citep{collins02}.  For prograde precession, the nodding period is $$P_n = {1\over2} {{P_p ~P_b}\over{P_p - P_b}}$$ where $P_p$ and $P_b$ are the precessional and binary periods, respectively.  The predicted nodding period, $P_n = 73.1$~d, is quite close to the observed $V/R$ modulation period of $69.3 \pm 0.2$~d discovered by \citet{pollmann08} and which is evident in the recent $V/R$ variations (Fig.~\ref{fig.vr}).  

A lingering difficulty for the precessing disk model is the fact that the intrinsic polarization angle has remained remarkably stable over the last decade \citep{mcdavid99,stefl09}, showing no evidence of changes as large as those seen in the interferometric data.  We suspect that this difference arises in the radial dependence of the disk shape.  The polarization measurements probe scattering radiation from the innermost part of the disk where the disk is probably coaligned with the stellar equator (the probable source of entering disk gas).  The tilt, on the other hand, is associated with vertical motions produced by the one-armed spiral oscillation that attain maximum amplitude at several stellar radii out into the disk.  Thus, we suggest that the sensitivity of the interferometric observations to disk emission at larger radii is the reason why the tilt oscillation is detected by interferometry and not detected by polarization measurements that reflect conditions in the inner disk.

\subsection{Nature of the Asymmetry in the Light Distribution}

 The non-zero closure phases we measure in $\zeta$ Tau imply the presence of an asymmetry in the disk and/or the star light distribution.  In general, the position angle of the peak of the skewed distribution lies within $\sim$$10^{\circ}$ of being perpendicular to the major axis of the disk.  However, because there is a fairly sharp transition between the ``bright" and ``dark'' sides of the disk, a small change in the angle of the asymmetry has a large effect on how the transition line intersects the outer edges of the disk and changes the brightness distribution there.

\begin{figure}
\begin{center}
   \scalebox{0.8}{\includegraphics{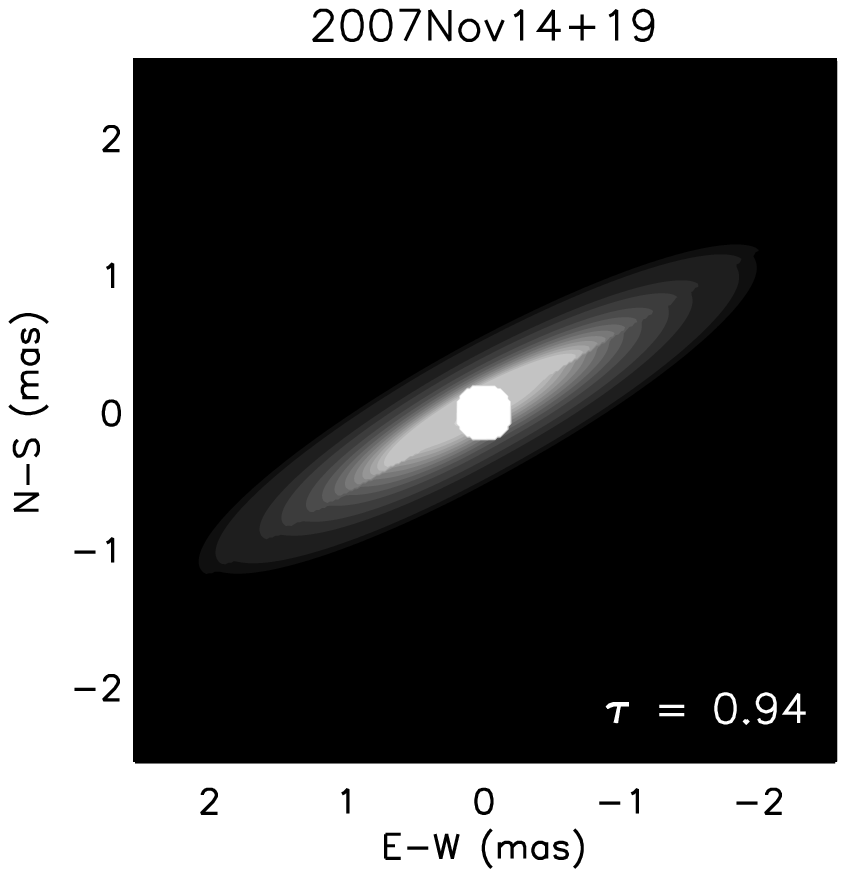}}
   \scalebox{0.8}{\includegraphics{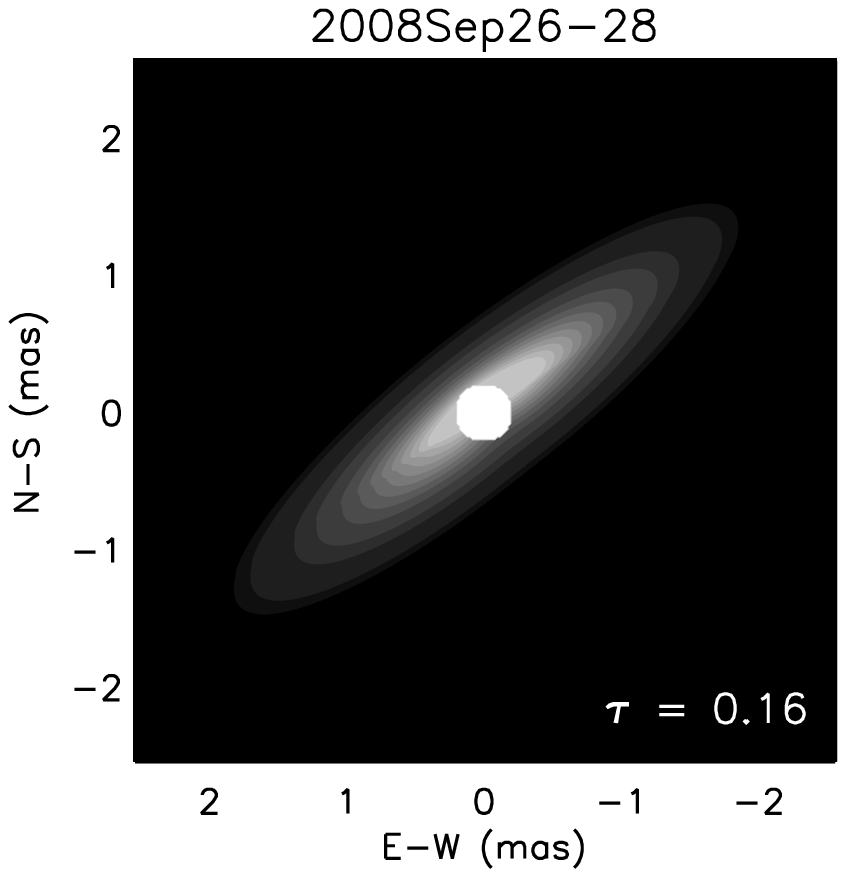}} \\
   \scalebox{0.8}{\includegraphics{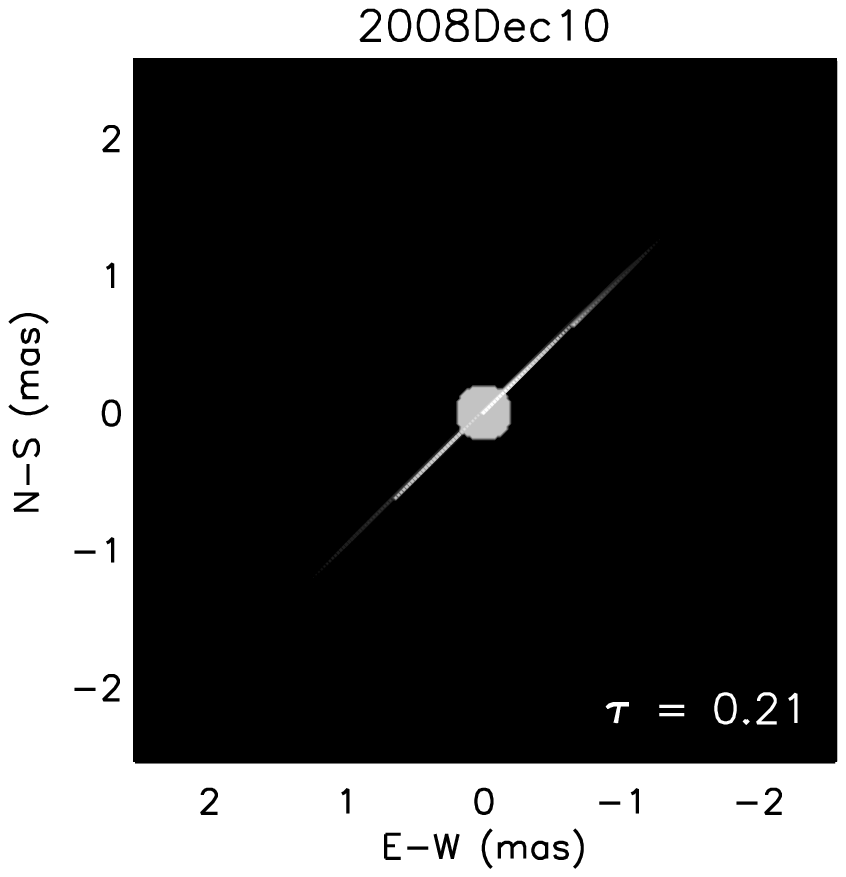}}
   \scalebox{0.8}{\includegraphics{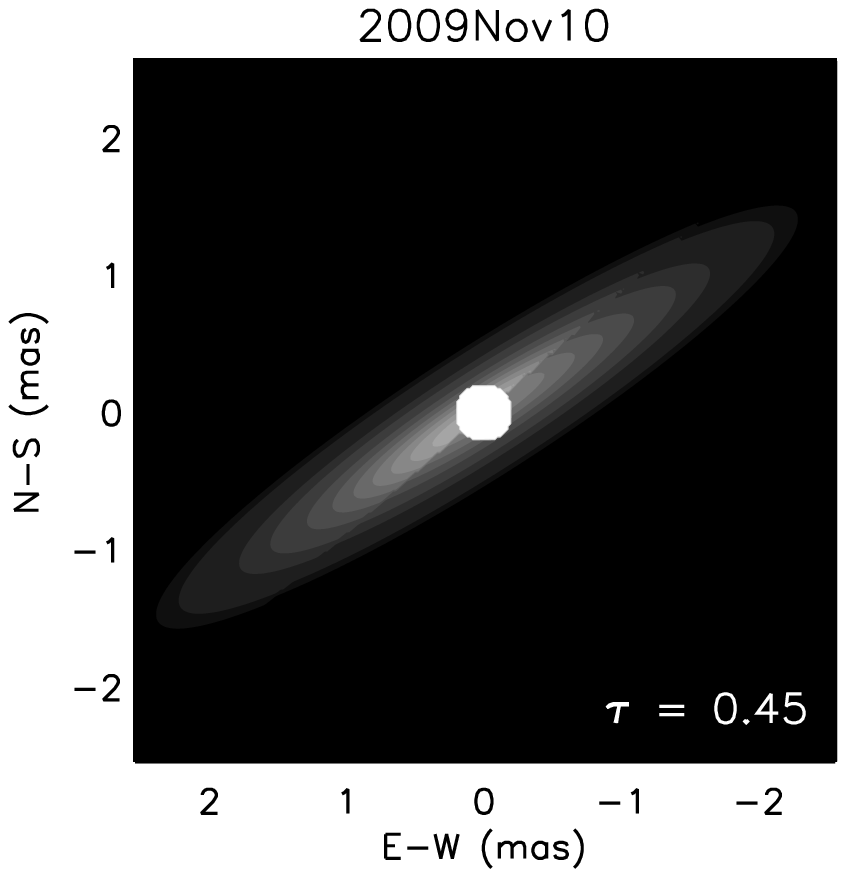}} \\
\end{center}
   \caption{Best-fit geometric models determined for $\zeta$ Tau during the epochs of the MIRC observations.  The spectroscopic $V/R$ phase $\tau$ is indicated in the bottom right of each panel.}
\label{fig.model}
\end{figure}

\begin{figure}
\begin{center}
  \rotatebox{90}{\scalebox{0.7}{\includegraphics{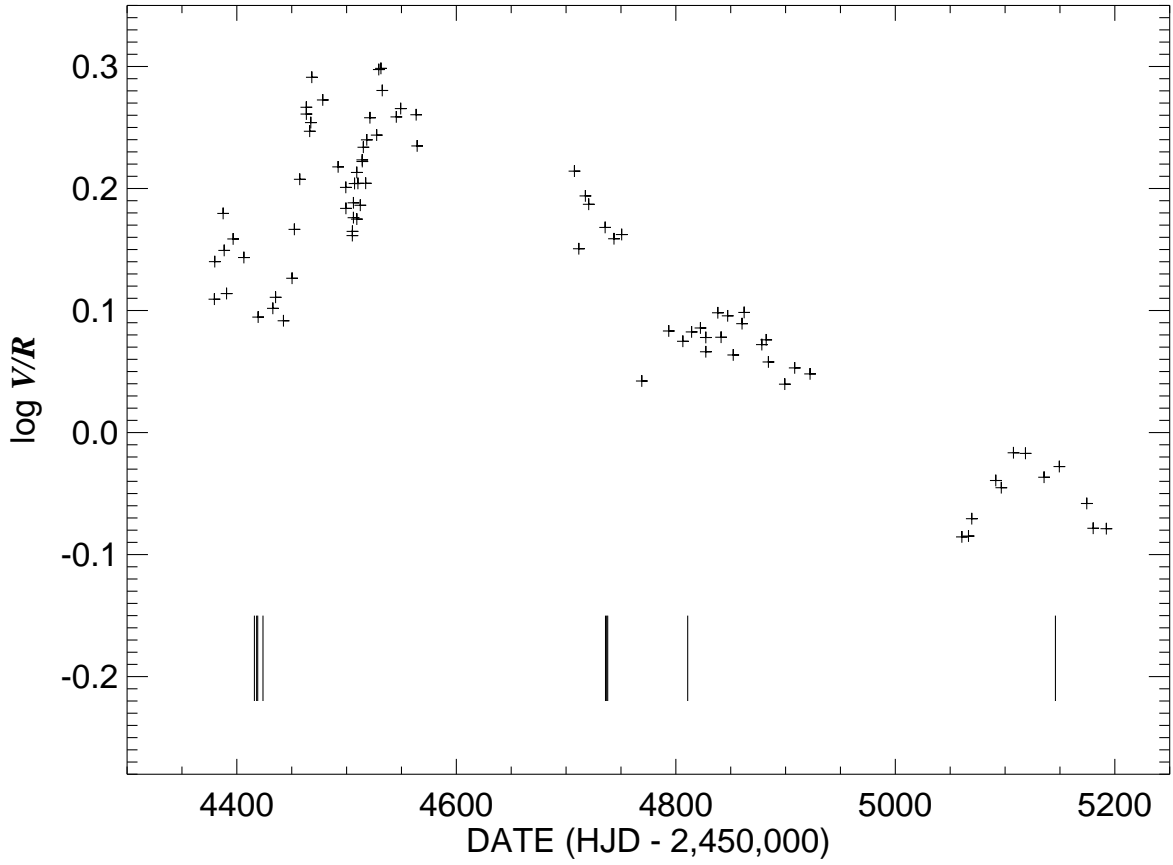}}}
\end{center}
\caption{The logarithm (base 10) of the $V/R$ ratio of the H$\alpha$ 
emission peaks as a function of time (heliocentric Julian date).
The vertical tick marks at bottom indicate the times of the 
MIRC observations.}
\label{fig.vr}
\end{figure}

\begin{figure}
\begin{center}
  \scalebox{0.80}{\includegraphics{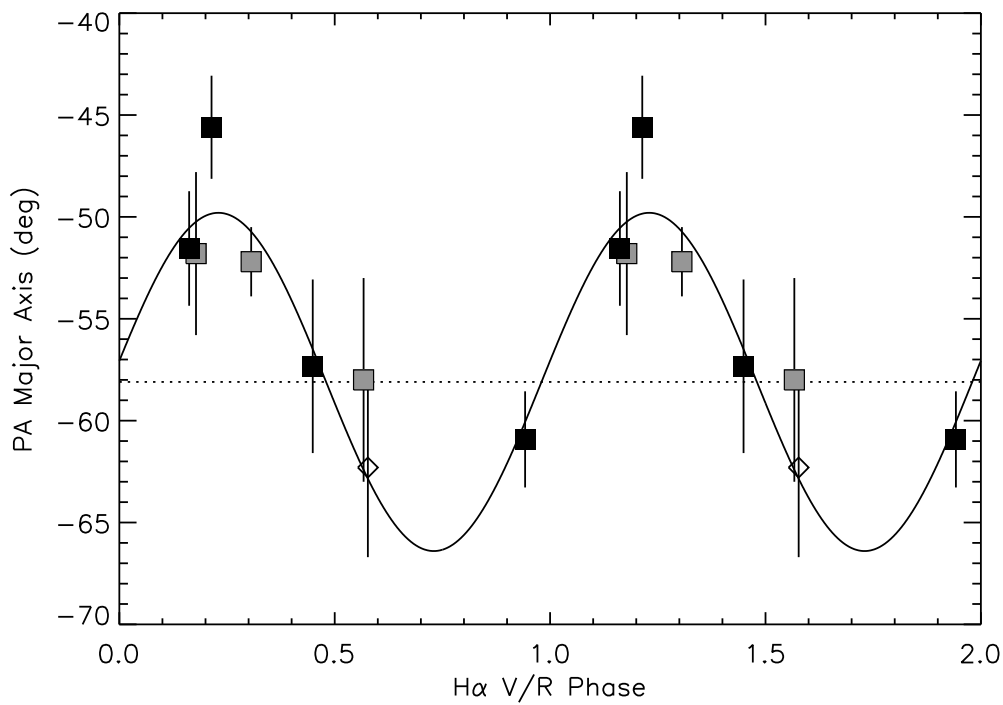}}
\end{center}
\caption{Disk long axis position angle as derived from interferometry 
plotted against $V/R$ phase $\tau$.  The measurements are repeated over
two cycles to emphasize phase continuity.  The filled black squares represent our MIRC observations (Table 3), the gray squares are the previously published near-IR measurements listed Table 5, and the open diamond indicates the H$\alpha$ result from Tycner et al.\ (2004).  The solid line shows a sinusoidal weighted fit of the variation, and the dotted line indicates the mean position angle determined independently from linear polarization observations.}
\label{fig.padisk}
\end{figure}

\begin{figure}
\begin{center}
   \rotatebox{90}{\scalebox{0.7}{\includegraphics{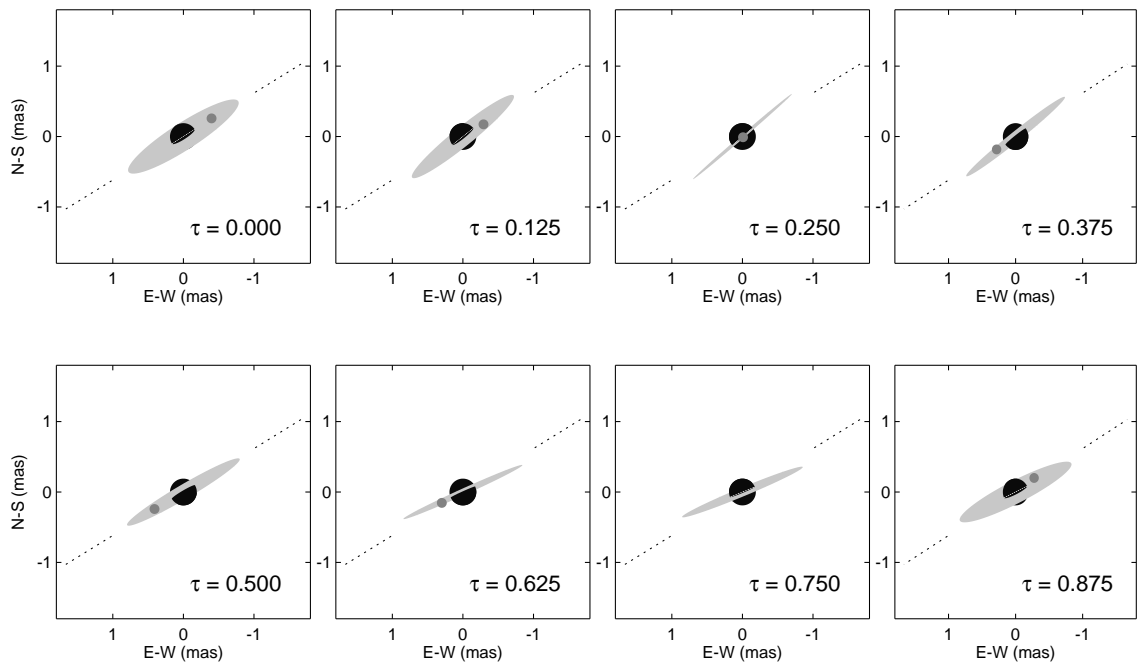}}}
\end{center}
\caption{Cartoon depiction of the disk precession variations as seen in 
the sky.  The Be star is shown as a black circle, the circumstellar disk as 
a gray ellipsoid, and the mean position angle of the long axis of the disk
(aligned with the stellar equator) as dotted lines extending
from the disk.  The panels show the change in the precession of the disk 
tilt and of the one-armed spiral density maximum ({\it small gray dot}) 
for eight steps in the precession cycle from $\tau=0$, the time of 
$V/R$ maximum.  The disk rotation and precession both advance around the 
Be star's spin axis (pointed to the south-west) so that the north-western
part of the disk approaches us.}
\label{fig.precess}
\end{figure}

From the model images in Figure~\ref{fig.model}, it appears that in 2007 Nov, the upper (north-eastern) half of the disk is brighter than the lower (south-western) half.  In 2008 Sep, the north-western side of the major axis appears brighter than the south-eastern side.   In 2009 Nov, the south-eastern side of the major axis is brighter than the north-western side.  These locations agree roughly with the quadrant where we expect to see the density enhancement based on the precession mode (gray circle in Fig.~\ref{fig.precess}), provided that the spiral arm extension covers a larger area of the disk than marked in the cartoon model.  A more sophisticated test of the one-armed spiral oscillation model would require adopting a more realistic brightness distribution for the density enhancement in addition to taking into account how the Be star is occulted by the disk.

In 2008 Dec, the disk appears very thin.  The skew parameters ($A_{\rm skew}$ = 1.0, $p$ = 0.001) indicate that the upper side of this thin disk is totally bright and the lower side is totally dark.  This suggests that the disk is tilted slightly toward the upper half of the star, similar to a phase between $\tau$=0.25 and $\tau$=0.375 displayed in the panels of Figure~\ref{fig.precess}.  Given the uncertainties in measuring the cycle length and the time of $V/R$ maximum, this is roughly consistent with our estimate of $\tau$=0.21 during this epoch.  The model image for 2008 Dec in Figure~\ref{fig.model} shows a slight enhancement in brightness toward the north-west side of the major axis.   However, we note that the skew axis $\phi_{\rm skew}=44\fdg11\pm0\fdg09$ deg is almost identical to the disk normal at a position angle $90^\circ+\phi_{\rm disk} = 44\fdg40\pm 0\fdg10$ deg, indicating that we cannot conclusively determine whether the brightness is skewed toward one side of the major axis.

If the binary companion of $\zeta$~Tau were bright enough to effect our observations, we would expect to see a periodic variation in the closure phases; the amplitude of the variation revealing the flux ratio and the periodicity corresponding to the separation on the sky \citep[e.g.][]{monnier07a}.  In Table~\ref{tab.binpos}, we list the predicted location of the binary companion at the times of the MIRC observations based on estimates of the orbital parameters.  To compute the separation ($\rho_{\rm bin}$) and position angle ($\phi_{\rm bin}$), we used the spectroscopic parameters derived by \citet{ruzdjak09} in their Table 2, Solution 2 ($P$ = 132.987 days, $K_1$ = 7.43 km\,s$^{-1}$, $e$ = 0, $\omega$ = 0).  Assuming that the binary orbit lies in the same plane as the disk and that the epoch of maximum radial velocity ($T_{\rm RV max}$ = 2,447,025.6 HJD) occurs when the companion is approaching from the north-west, we adopted an orbital inclination of $i = 92\fdg8$ and a position angle of the line of nodes of $\Omega = -58\fdg0$.  If we assume that the B2 IIIpe primary has a mass of 11.2 $M_\odot$ \citep{gies07}, these orbital parameters yield a secondary with a mass of 0.94 $M_\odot$.  Using Kepler's Third Law, these mass estimates imply a semi-major axis for the binary orbit of 1.17 AU, or $a$ = 9.17 mas at a distance of 128 pc.  If the companion is a main sequence star, the secondary mass corresponds roughly to a G4 spectral type with an absolute magnitude of $M_V$ = 5.0, $V-K$ = 1.5, and $K-H$ = 0.1 \citep{AQ}.  Using the distance modulus implied by the Hipparcos parallax, this corresponds to an apparent $H$-band magnitude of 9.0.  Compared with the 2MASS magnitude of $\zeta$ Tau ($H$=3.0), this corresponds to a magnitude difference of $\Delta H$ = 5.9.  

It could be argued that we measure a small periodic amplitude variation in the closure phases plotted in Figure~\ref{fig.t3}.  However, the data quality is not sufficient to compute a full binary fit.  Adding a binary component to our skewed elliptical Gaussian models at a fixed flux ratio and a position given by the values in Table~\ref{tab.binpos}, does not improve the $\chi^2$ significantly.  Even though we do not detect the companion, the variation in the closure phase residuals allows us to put an upper limit on the binary flux ratio.  The maximum deviation from the best fit skewed disk model ranges from $2\fdg2$ in 2007~Nov to $7\fdg1$ in 2009~Dec.  If this variation is due only to the signature of a binary companion, then we can place a lower limit on the $H$-band magnitude difference between $\zeta$~Tau and the companion of 5.3 mag and 3.3 mag, respectively.  These values are consistent with the expected magnitude difference of $\Delta H$ = 5.9 estimated from the orbital parameters we assumed.
The observed magnitude limit has interesting implications for the possibility 
of a hot subdwarf companion.  If we take the effective temperature and radius for the hot subdwarfs in 
$\phi$~Per \citep{gies98} and FY~CMa \citep{peters08} and assume a Planck 
flux distribution, then we would predict magnitude
differences of $\Delta H$ = 2.5 ($\phi$ Per subdwarf) and 4.4 (FY CMa subdwarf),
which would probably be detectable.  So if the companion is a subdwarf, it is probably
cooler and/or smaller than the subdwarfs detected in the other two cases.

\section{Conclusions}
\label{sect.summary}

We obtained four epochs of interferometric measurements on the Be star $\zeta$~Tau using the MIRC beam combiner at the CHARA Array.  By fitting the disk with a skewed elliptical Gaussian model, we determine a full width at half maximum size of the major axis of $\sim$ 1.8 mas and estimate that the central Be star contributes 55\% of the light in the $H$-band.  Combining our results with previous interferometric measurements, we observe a change in the position angle of the disk over time.  A correlation between the position angle and the $V/R$ phase of the H$\alpha$ emission line variation suggests that the tilt of the disk around $\zeta$~Tau is precessing.  The tilt could be generated by vertical motions of the gas caused by the spiral density enhancement \citep{ogilvie08,oktariani09} as it moves through the disk.  We also measure an asymmetry in the light distribution of the disk that roughly corresponds to the expected location of the density enhancement in the spiral oscillation model.

We plan to continue monitoring changes in the structure and orientation of the disk of $\zeta$ Tau with future observations at the CHARA Array.  Ultimately, these observations can be used to test predictions from one-armed spiral oscillation models (e.g. Berio et al. 1999; Meilland et al. 2007; Jones et al. 2008; Carciofi et al. 2009).

\acknowledgments

We thank Christian Buil, Benjamin Mauclaire, Ernst Pollmann, and the other
observers who contributed to the BeSS database of $\zeta$ Tau spectroscopy. 
We also thank K.\ Bjorkman, E.\ Hesselbach, E.\ Grundstrom, and V.\ McSwain for 
providing 
additional spectra contemporaneous with the MIRC observations.  We thank the 
anonymous referee for providing comments that helped improve the paper.
DRG acknowledges support for this work provided by the National Science
Foundation under grant AST-0606861. JDM, XC, and MZ acknowledge funding from
Univ. Michigan and the National Science Foundation
(AST-0352723, AST-0707927, AST-0807577). STR acknowledges partial support by 
NASA Grant NNH09AK731.  
Operational funding for the CHARA Array is provided by the GSU College
of Arts and Sciences, by the National Science Foundation through grants
AST-0606958 and AST-0908253, by the W.M. Keck Foundation, and by the NASA 
Exoplanet Science Institute.
We thank the Mount Wilson Institute for providing infrastructure support at
Mount Wilson Observatory. The CHARA Array, operated by Georgia State
University, was built with funding provided by the National Science
Foundation, Georgia State University, the W. M. Keck Foundation, and the
David and Lucile Packard Foundation.
We gratefully acknowledge all of this support.
This research has made use of the SIMBAD database and the VizieR catalog 
service operated at CDS, Strasbourg, France.

\clearpage

\clearpage

\begin{deluxetable}{lll}
\tablewidth{0pt}
\tablecaption{CHARA-MIRC Observing Log for $\zeta$ Tau}
\tablehead{\colhead{UT Date} & \colhead{Configuration} & \colhead{Calibrators}}
\startdata
2007 Nov 11  &  S2-E2-W1-W2  &  $\zeta$ Per \\
2007 Nov 13  &  S2-E2-W1-W2  &  $\zeta$ Per \\
2007 Nov 14  &  S2-E2-W1-W2  &  $\sigma$ Cyg, $\zeta$ Per \\
2007 Nov 19  &  S1-E1-W1-W2  &  $\zeta$ Per, 10 Aur \\
2008 Sep 26  &  S1-E1-W1-W2  &  $\zeta$ Per, $\theta$ Gem \\
2008 Sep 27  &  S1-E1-W1-W2  &  $\zeta$ Per, $\theta$ Gem \\ 
2008 Sep 28  &  S1-E1-W1-W2  &  $\zeta$ Per \\ 
2008 Dec 10  &  S1-E1-W1-W2  &  $\zeta$ Per, 15 LMi \\
2009 Nov 10  &  S1-E1-W1-W2  &  HR 485, $\zeta$ Per, 71 Ori, 79 Cnc
\label{tab.log}
\enddata
\end{deluxetable}

\begin{deluxetable}{lllllrcrcl}
\tabletypesize{\scriptsize}
\tablewidth{0pt}
\tablecaption{Calibrator Diameters}
\tablehead{\colhead{Calibrator} & \colhead{HD Number} & \colhead{SpT} & \colhead{$V$} & \colhead{$H$} & \colhead{$T_{\rm eff}$} & \colhead{Ref} & \colhead{$E(B-V)$} &  \colhead{$R$} & \colhead{LD Diameter}  \\ \colhead{} & \colhead{} & \colhead{} & \colhead{(mag)} & \colhead{(mag)} & \colhead{(K)} & \colhead{} & \colhead{(mag)}  & \colhead{}  & \colhead{(mas)}}
\startdata
HR 485       & HD 10348  & K0 III & 5.97  & 3.83   &  4885 & 1 & 0.046 $\pm$ 0.031 & 3.1             & 0.879 $\pm$ 0.046 \\
$\zeta$ Per  & HD 24398  & B1 Iab & 2.87  & 2.62   & 21950 & 2 & 0.355 $\pm$ 0.012 & 2.88 $\pm$ 0.09 & 0.645 $\pm$ 0.026 \\
10 Aur       & HD 32630  & B3 V   & 3.16  & 3.76   & 16600 & 3 & 0.024 $\pm$ 0.007 & 3.1             & 0.444 $\pm$ 0.012 \\
71 Ori       & HD 43042  & F6 V   & 5.20  & 3.83   &  6485 & 4 & 0     $\pm$ 0.007 & 3.1             & 0.597 $\pm$ 0.021 \\
$\theta$ Gem & HD 50019  & A3 III & 3.60  & 3.23   &  8300 & 5 & 0.033 $\pm$ 0.009 & 3.1             & 0.796 $\pm$ 0.022 \\
79 Cnc       & HD 78715  & G5 III & 6.0   & 4.09   &  5050 & 6 & 0     $\pm$ 0.007 & 3.1             & 0.747 $\pm$ 0.062 \\
15 LMi       & HD 84737  & G0 V   & 5.08  & 3.61   &  5830 & 7 & 0.046 $\pm$ 0.024 & 3.1             & 0.849 $\pm$ 0.022 \\
$\sigma$ Cyg & HD 202850 & B9 Iab & 4.25  & 3.86   & 11000 & 8 & 0.237 $\pm$ 0.009 & 3.1             & 0.574 $\pm$ 0.017 
\label{tab.cal}
\enddata
\tablerefs{(1) Hekker \& Mel\'{e}ndez 2007; (2) Huang \& Gies 2008; (3) Lyubimkov et al.\ 2002; (4) Lambert \& Reddy 2004; (5) Malagnini et al.\ 1982; (6) de Laverny et al. 2003; (7) Ram\'{i}rez et al.\ 2007; (8) Markova \& Puls 2008}
\end{deluxetable}

\clearpage

\begin{deluxetable}{lccccccc}
\rotate
\tabletypesize{\footnotesize}
\tablewidth{0pt}
\tablecaption{Disk Modeled as an Elliptical Gaussian Modulated by a Sinusoid}
\tablehead{\colhead{Parameter} & \colhead{2007Nov11,13} & \colhead{2007Nov14} & \colhead{2007Nov19} & \colhead{2007Nov14,19} & \colhead{2008Sep26-28} & \colhead{2008Dec10} & \colhead{2009Nov10} \\	    
           \colhead{}          & \colhead{S2E2W1W2}     & \colhead{S2E2W1W2}  & \colhead{S1E1W1W2}  &  \colhead{S2E2W1W2}    & \colhead{S1E1W1W2}      & \colhead{S1E1W1W2}  & \colhead{S1E1W1W2} \\
           \colhead{}          & \colhead{}             & \colhead{}          & \colhead{}          &  \colhead{S1E1W1W2}    & \colhead{} & \colhead{} & \colhead{}} 
\startdata
$f_{\rm disk}$              & 0.5948$\pm$0.0073 & 0.5464$\pm$0.0078 & 0.5036$\pm$0.0065 & 0.5114$\pm$0.0065 & 0.4717$\pm$0.0060 & 0.4669$\pm$0.0243 & 0.3480$\pm$0.0051 \\  
$f_{\rm star}$              & 0.4052$\pm$0.0073 & 0.4536$\pm$0.0078 & 0.4964$\pm$0.0065 & 0.4886$\pm$0.0065 & 0.5283$\pm$0.0060 & 0.5331$\pm$0.0243 & 0.6520$\pm$0.0051 \\ 
$\theta_{\rm maj}$ (mas)    & 1.662$\pm$0.042   & 1.527$\pm$0.045   & 1.410$\pm$0.025   & 1.599$\pm$0.036   & 1.633$\pm$0.031   & 1.116$\pm$0.061   & 2.057$\pm$0.060   \\ 
$\theta_{\rm min}$ (mas)    & 0.691$\pm$0.027   & 0.598$\pm$0.047   & 0.308$\pm$0.015   & 0.331$\pm$0.023   & 0.401$\pm$0.010   & 0.020             & 0.346$\pm$0.016   \\ 
$\theta_{\rm min}/\theta_{\rm maj}$ & 0.416$\pm$0.019   & 0.392$\pm$0.033   & 0.218$\pm$0.011   & 0.207$\pm$0.015   & 0.246$\pm$0.008   & 0.018$\pm$0.001   & 0.168$\pm$0.009   \\ 
$\phi_{\rm disk} (^\circ)$  & $-$61.82$\pm$0.86 & $-$62.07$\pm$0.75 & $-$58.83$\pm$0.71 & $-$60.92$\pm$0.66 & $-$51.55$\pm$0.51 & $-$45.60$\pm$0.10 & $-$57.33$\pm$0.86   \\  
$A_{\rm skew}$              & 0.92$\pm$0.11     & 0.35$\pm$0.70     & 0.34$\pm$0.02     & 0.40$\pm$0.08     & 0.49$\pm$0.09     & 1.00              & 0.21$\pm$0.02     \\ 
$\phi_{\rm skew} (^\circ)$  & 29.49$\pm$ 0.65   & 29.03$\pm$ 0.22   & 32.42$\pm$ 0.37   & 30.44$\pm$ 0.62   & 32.88$\pm$ 0.53   & 44.11$\pm$ 0.09   & 41.28$\pm$ 1.11   \\  
$p$                         & 1.00              & 0.05$\pm$1.35     & 0.001             & 0.16$\pm$0.09     & 0.52$\pm$0.11     & 0.001             & 0.001     	      \\ 
$\chi_{\nu}^2$(all)         & 2.25             & 1.11             & 1.12             & 2.16             & 1.27             & 1.16             & 1.33	      \\ 
$\chi_{\nu}^2 (V^2)$        & 3.40             & 1.31             & 1.09             & 2.46             & 1.09             & 1.74             & 1.47	      \\ 
$\chi_{\nu}^2$ (T3)         & 0.64             & 0.94             & 1.33             & 1.84             & 1.61             & 0.58             & 1.16             \\ 
HJD $-$ 2,400,000       &  54417.0          &  54419.0           &  54423.8          &  54421.4            & 54736.9          & 54810.8           & 55145.9               \\ 
$\tau$ ($V/R$ Phase)       &  \nodata          &  \nodata           &  \nodata          &  0.942              & 0.162            & 0.214             & 0.449  
\label{tab.diskfit}
\enddata
\tablecomments{Values without error bars indicate that the fitting routine reached the limit of the search range.}
\end{deluxetable}

\begin{deluxetable}{llllll}
\tabletypesize{\footnotesize}
\tablewidth{0pt}
\tablecaption{Systematic Errors in Model Parameters Estimated Through a Monte Carlo Analysis}
\tablehead{\colhead{Parameter}    & \colhead{2007Nov11-14} & \colhead{2007Nov14,19} & \colhead{2008Sep26-28} & \colhead{2008Dec10} & \colhead{2009Nov10} \\
           \colhead{}             & \colhead{S2E2W1W2}     & \colhead{S1E1W1W2}     & \colhead{S1E1W1W2}     & \colhead{S1E1W1W2}  & \colhead{S1E1W1W2} \\
           \colhead{}             & \colhead{}             & \colhead{S2E2W1W2}     & \colhead{}             & \colhead{}          & \colhead{}} 
\startdata
$f_{\rm disk}$            & $\pm$0.027  & $\pm$0.020  & $\pm$0.030   & $\pm$0.091  & $\pm$0.027 \\
$\theta_{\rm maj}$ (mas)  & $\pm$0.40   & $\pm$0.12   & $\pm$0.18    & $\pm$0.15  & $\pm$0.45  \\
$\theta_{\rm min}$ (mas)  & $\pm$0.15   & $\pm$0.074  & $\pm$0.047   & $\pm$0.080 & $\pm$0.073 \\
$\phi_{\rm disk} (^\circ)$ & $\pm$13.9   & $\pm$2.4    & $\pm$2.8     & $\pm$2.5   & $\pm$4.3
\label{tab.syserr}
\enddata
\end{deluxetable}

\begin{deluxetable}{lccccc}
\tabletypesize{\footnotesize}
\tablewidth{0pt}
\tablecaption{Previous Interferometric Results}
\tablehead{\colhead{Parameter} & \colhead{1992.82} & \colhead{1999.16}  & \colhead{2005.93} & \colhead{2006.95} & \colhead{2008.80} }
\startdata
$f_{\rm star}$                       &  0.30          & 0.814$\pm$0.012 & 0.414$\pm$0.029 & \nodata      & \nodata         \\
$\theta_{\rm maj}$ (mas)             &  4.53$\pm$0.52 & 3.14$\pm$0.21   & 1.79$\pm$0.07   & \nodata      & \nodata         \\	
$\theta_{\rm min}/\theta_{\rm maj}$  &  0.28$\pm$0.28 & 0.310$\pm$0.072 & 0.09$\pm$0.22   & \nodata      & \nodata         \\
$\phi_{\rm disk}$ ($^\circ$)         &  $-$58$\pm$4   & $-$62.3$\pm$4.4 & $-$52.2$\pm$1.7 & $-$58$\pm$5  & $-$51.8$\pm$4.0 \\	
Observatory                          &  MkIII         &  NPOI           & CHARA-Classic   & VLTI-AMBER   & CHARA-Classic   \\
Filter                               &  H$\alpha$     &  H$\alpha$      & $K'$            & $K$          &  $K'$           \\
HJD $-$ 2,400,000                     &  48915.8       & 51238.8         & 53709.3         & 54081.7      & 54758.9         \\
$\tau$ ($V/R$ Phase)                 &  \nodata       & 0.577           & 0.306           & 0.567        &  0.178          \\
Reference                            &  1             & 2               & 3               & 4, 5          &  6              
\label{tab.compare}
\enddata
\tablerefs{(1) Quirrenbach et al. 1997; (2) Tycner et al. 2004; (3) Gies et al. 2007; (4) \v{S}tefl et al. 2009; (5) Carciofi et al. 2009; (6) Touhami et al., in preparation}
\end{deluxetable}

\begin{deluxetable}{llllll}
\tablewidth{0pt}
\tablecaption{Predicted Location of the Binary Companion During the MIRC Observations}
\tablehead{  & \colhead{2007Nov14} & \colhead{2007Nov19} & \colhead{2008Sep26-28} & \colhead{2008Dec10} & \colhead{2009Nov10}}
\startdata    
HJD $-$ 2,400,000         &  54419.0          &  54423.8          & 54736.9              & 54810.8           & 55145.9           \\ 
$\rho_{\rm bin}$ (mas)     &  7.59             &  6.24             & 9.13                 & 8.86              & 8.51              \\
$\phi_{\rm bin}$ ($^\circ$) &  120.1            &  119.0           & 302.3               & 121.3            & 300.9 
\label{tab.binpos}
\enddata
\end{deluxetable}

\clearpage



\clearpage

\clearpage

\clearpage

\clearpage

\clearpage

\end{document}